\documentclass[a4paper,twocolumn,11pt,accepted=2023-08-17]{quantumarticle}
\pdfoutput=1

\usepackage{bbm, amsthm, bm,textcomp, nicefrac,geometry,ragged2e, multirow}
\usepackage[dvipsnames]{xcolor}
\usepackage{float}
\usepackage[bbgreekl]{mathbbol}
\usepackage{graphicx,epstopdf,color,verbatim,enumitem}
\usepackage[numbers, sort&compress]{natbib} 
\usepackage[normalem]{ulem}
\usepackage{pbox,array}
\usepackage{mathrsfs}
\usepackage{amssymb}
\usepackage{stackrel}
\usepackage{amsmath}
\usepackage{physics}
\makeatletter
\usepackage[T1]{fontenc}
\usepackage[english]{babel}
\addto\captionsenglish{\renewcommand{\figurename}{FIG.}}
\usepackage{dcolumn}    
\usepackage{ifpdf}
\usepackage{lineno,amsfonts}
\usepackage{graphicx}   
\usepackage{bm}         
\usepackage{bbm}
\usepackage{mathrsfs}
\usepackage{upgreek}
\usepackage{mathtools}
\usepackage{epstopdf}
\usepackage{setspace}
\usepackage{hyperref}
\usepackage{float}
\usepackage{ifpdf}
\usepackage{qcircuit}
\usepackage{dsfont}

\newcommand{\expec}[1]{\langle{#1}\rangle}
\newcommand{\ez}{\ensuremath{\varepsilon_0}}

\begin{document}

\title{Simulating gauge theories with variational quantum eigensolvers in superconducting microwave cavities}

\author{Jinglei Zhang}
\thanks{These authors contributed equally to this work}
\affiliation{Institute for Quantum Computing, University of Waterloo, Waterloo, Ontario N2L 3G1, Canada}
\affiliation{Department of Physics and Astronomy, University of Waterloo, Waterloo, Ontario N2L 3G1, Canada}
\orcid{0000-0002-3100-7921}
\author{Ryan Ferguson}
\thanks{These authors contributed equally to this work}
\affiliation{Institute for Quantum Computing, University of Waterloo, Waterloo, Ontario N2L 3G1, Canada}
\affiliation{Department of Physics and Astronomy, University of Waterloo, Waterloo, Ontario N2L 3G1, Canada}
\author{Stefan K\"uhn} 
\affiliation{Computation-based Science and Technology Research Center,The Cyprus Institute, 20 Kavafi Street, 2121 Nicosia, Cyprus}
\author{Jan F. Haase}
\affiliation{Institute for Quantum Computing, University of Waterloo, Waterloo, Ontario N2L 3G1, Canada}
\affiliation{Department of Physics and Astronomy, University of Waterloo, Waterloo, Ontario N2L 3G1, Canada}
\affiliation{Institute of Theoretical Physics and IQST, Universit{\"a}t Ulm, Albert-Einstein-Allee 11, D-89069 Ulm, Germany}
\orcid{0000-0003-1126-8216}
\author{C.M.~Wilson}
\affiliation{Institute for Quantum Computing, University of Waterloo, Waterloo, Ontario N2L 3G1, Canada}
\affiliation{Department of Electrical and Computer Engineering, University of Waterloo, Waterloo, Ontario N2L 3G1, Canada}
\author{Karl Jansen} 
\affiliation{NIC,  DESY  Zeuthen,  Platanenallee  6,  15738  Zeuthen,  Germany}
\author{Christine A. Muschik}
\affiliation{Institute for Quantum Computing, University of Waterloo, Waterloo, Ontario N2L 3G1, Canada}
\affiliation{Department of Physics and Astronomy, University of Waterloo, Waterloo, Ontario N2L 3G1, Canada}
\affiliation{Perimeter Institute for Theoretical Physics, Waterloo, Ontario N2L 2Y5, Canada}
\orcid{0000-0002-4599-5107}

\begin{abstract}
Quantum-enhanced computing methods are promising candidates to solve currently intractable problems. We consider here a variational quantum eigensolver (VQE), that delegates costly state preparations and measurements to quantum hardware, while classical optimization techniques guide the quantum hardware to create a desired target state. In this work, we propose a bosonic VQE using superconducting microwave cavities, overcoming the typical restriction of a small Hilbert space when the VQE is qubit based. The considered platform allows for strong nonlinearities between photon modes, which are highly customisable and can be tuned in situ, i.e. during running experiments. Our proposal hence allows for the realization of a wide range of bosonic ansatz states, and is therefore especially useful when simulating models involving degrees of freedom that cannot be simply mapped to qubits, such as gauge theories, that include components which require infinite-dimensional Hilbert spaces. We thus propose to experimentally apply this bosonic VQE to the U(1) Higgs model including a topological term, which in general introduces a sign problem in the model, making it intractable with conventional Monte Carlo methods.
 \end{abstract}
\newline
\\
 
 \maketitle
 
 \tableofcontents
 
 \section{Introduction \label{sec:intro}}

Quantum computers have the potential to solve currently intractable problems in fundamental and applied sciences. In particular, hybrid quantum-classical approaches bring interesting applications closer in reach by lowering the requirements for quantum hardware. In this paper, we pursue this direction by considering a variational quantum eigensolver (VQE)~\cite{farhi2014quantum, mcclean_theory_2016, preskill2018quantum, cerezo_variational_2021}. This algorithm takes advantage of a quantum processor to efficiently generate a parametrized ansatz state. An appropriate cost function is measured on the state and fed into a classical processor. The classical processor uses this information to minimize the cost function by optimizing the parameters that control the ansatz state (see App.~\ref{app:intro_vqs} for more details). VQEs exploit the quantum processor to efficiently evaluate a cost function that is hard to calculate classically, while the variational nature of the optimization algorithm ensures resilience to certain types of errors and lowers requirements for the quantum hardware. This approach has been shown to be extremely successful~\cite{muschik_u1_2017,Kandala2017,Kokail2018,Klco2018}. 

While VQEs have been widely implemented with qubit-based quantum hardware~\cite{mezzacapo_non-abelian_2015, yang_observation_2020, klco_su2_2020, atas_su2_2021, lumia_two-dimensional_2022, zhou_thermalization_2022}, we propose in this work a \textit{bosonic} VQE using superconducting microwave cavities. Our approach is especially useful when simulating models involving degrees of freedom which are not well-described as qubits. Important examples are gauge theories that include quantum fields to be simulated, which are defined in infinite-dimensional Hilbert spaces. Photonic systems are apt candidates for simulating such models, but the advantage of the large photonic Hilbert space usually comes with the disadvantage that strong nonlinear interactions are not easily available. By contrast, our proposal builds on recent advances in superconducting microwave cavities that allow for strong nonlinearities between photon modes~\cite{chang_generating_2018, alaeian_creating_2019, chang_observation_2020, hung_quantum_2021, vrajitoarea_quantum_2020}. In superconducting microwave cavities, Josephson junctions are an important source of strong nonlinear dynamics: through parametric processes it is possible to easily select and enhance a variety of interactions between the cavity modes. This flexibility in the available interactions is an important resource in a VQE, since they determine the form of the ansatz state and are fundamental in the success of the algorithm. In microwave cavities, nonlinearities have been demonstrated to be highly customisable, and can be programmed in situ, i.e. during running experiments, by changing classical microwave fields generated at room temperature. Using these programmable microwave fields for realizing VQE schemes allows, therefore, for the realization of a wide range of bosonic ansatz states.

As a concrete example, we propose in this paper to simulate topological terms in gauge theories. Topological terms are particularly challenging for classical numerical studies of lattice gauge theories since they give rise to a sign problem~\cite{Sulejmanpasic:2019ytl,Gattringer2018, Goschl:2018uma}. However, they play an important and fascinating role in quantum field theory. It would be very interesting and important to determine the amount of CP violation emerging from QCD itself to shed light on the the strong CP problem~\cite{Dar:2000tn} and, in turn, on physics beyond the standard model~\cite{Pospelov:2005pr}. In analogy to QCD, also in the electroweak sector of the standard model, a CP violating topological term can be introduced which can lead to enhanced baryon number violating processes~\cite{Cohen:1993nk}. 
This mechanism can thus provide one of the basic ingredients for generating the matter-antimatter asymmetry of the universe. A topological term can already be realized in one spatial dimension using a compact formulation of a U(1) gauge theory. This allows us to study the effect of such a term, the strength of which is given by the parameter $\theta$, in a much simpler setup than the full electroweak sector and constitutes a very first step towards addressing the complete gauge Higgs sector of the standard model. In addition to this perspective within the high-energy physics context, a gauge-Higgs model with a topological term is also closely related to interesting models in condensed matter physics~\cite{Goschl:2018uma,Ichinose2014,Komargodski2019}. We note that compact 1+1 dimensional U(1) gauge theories in the presence of a nonzero topological term are also of interest by themselves. In particular, the topological term can lead to new phenomena~\cite{Coleman1976,Adam1997,Byrnes2002,Buyens2017,Zache2019,Funcke2019}, and an enriched phase structure of the model~\cite{Anosova:2019quw,Sulejmanpasic:2019ytl,Komargodski2019}. The rich physical content found in this model can in general \textit{not} be explored with conventional Markov Chain Monte Carlo (MCMC) lattice methods because the topological term induces a sign problem in Euclidean time (it has been realized though that in the special case of 1+1 dimensions the model can be written in a dual formulation which is sign-problem free~\cite{Sulejmanpasic:2019ytl,goschl2018dual}). Importantly, the topological term studied here can be extended to three spatial dimensions~\cite{kan_investigating_2021}, showing a promising path to explore new physics with quantum simulators.

More specifically, we present an experimental proposal for simulating the U(1) Abelian Higgs lattice gauge theory in one spatial dimension with a topological term and open boundary conditions~\cite{Fradkin1979,Jones1979,Gonzalez-Cuadra2017}. We show how to use a VQE in order to study the phase diagram of the model. We also explore the physics of the model using matrix product states (MPS)~\cite{Verstraete2008,Schollwoeck2011,Orus2014a}. This allows us to diagonalize the Hamiltonian on lattices of various sizes and various truncations of the Higgs field's Hilbert space.  We find that, despite the finite lattice size and truncated operators, we are able to capture the physics of the main features of the phase diagram from analytic results, and conclude that it can be simulated on near-term quantum devices.

The rest of the paper is organized as follows. In Sec.~\ref{sec:experiment} we introduce the gauge theory under consideration, and explain how it can be mapped to a model suitable for simulations with parametric cavities. We then describe the experimental system, and explain how the experimental resources can be adjusted to perform a bosonic VQE for the U(1) Higgs model with topological term. In Sec.~\ref{sec:vqe} we discuss in detail the VQE protocol and results. To put theses results into context, we summarize the main features of the phase structure of the U(1) Higgs model in Sec.~\ref{sec:phase_diagram}. Subsequently we numerically demonstrate, using MPS, that we can capture the relevant physics of the model, even for the limited system sizes that can be simulated on current quantum hardware, and we systematically explore finite-size effects in Sec.~\ref{sec:mps}. We summarize our findings and offer some possible perspectives in Sec.~\ref{sec:conclusion}.

 \section{VQE with parametric cavities \label{sec:experiment}}

 In this section, we discuss how the physics of a gauge theory can be studied using a superconducting microwave cavity, using the one dimensional Higgs model with topological term as concrete example. As outlined in the introduction, we propose a VQE approach, which forgoes the need to implement the complicated interactions that appear in the Higgs model on the quantum simulator. Instead, the VQE protocol exploits a set of resource Hamiltonians that can be realized on a given platform, and allows for the preparation of an output state that approximates the targeted ground state using the limited set of resource interactions available (see App.~\ref{app:intro_vqs} for a more detailed discussion about VQEs). We target quantum simulations with parametric microwave cavities~\cite{chang_generating_2018, chang_observation_2020,hung_quantum_2021}, which have not been used for VQE so far, but are a promising candidate system as we explain below. 
 In particular, we first introduce in Sec.~\ref{subsec:Hamiltonian} the Hamiltonian of the model, which involves both scalar Higgs and gauge fields. We then describe the microwave platform and the features that make it suitable for our simulation in Sec.~\ref{sec:platform}. In Sec.~\ref{sec:HOBM}, we give the mapping from the fields of the original model to the photonic modes experimentally available. We present the specific resource Hamiltonian that can be engineered in the system, and the measurement scheme necessary to run the VQE in Sec.~\ref{subsec:resource_ham} and Sec.~\ref{subsec:measurements} respectively. We conclude by discussing the experimental imperfections and their role for the proposed simulation in Sec.~\ref{subsec:errors}.

 \subsection{The Hamiltonian\label{subsec:Hamiltonian}}

  Contrary to conventional lattice methods, for the purpose of quantum simulations, it is advantageous to work with a Hamiltonian lattice formulation. Here, we describe the U(1) Higgs lattice Hamiltonian with a topological term in one spatial dimension with open boundary conditions. 
  
  The Higgs fields are defined on the lattice sites, and the gauge fields are defined on the links between the lattice sites. In particular, the Higgs field on site $n$ is $\hat{\phi}_n$ and it has a canonically conjugate operator $\hat{Q}_n$, called the charge operator. Considering the case of fixed length of the Higgs field, these operators satisfy the commutation relation~\cite{Fradkin1979,Jones1979,Gonzalez-Cuadra2017}
 \begin{equation} 
     \label{eq:first_comm}
     [\hat{Q}_n, \hat{\phi}_{n'}^\dagger] = \delta_{n,n'}\hat{\phi}_n^\dagger,
 \end{equation}
 and as a result the Higgs field $\hat{\phi}_{n}$ acts as a lowering operator for the eigenstates of $\hat{Q}_n$: 
 \begin{subequations}
     \begin{align} 
     \hat{Q}_n\ket{Q}_n &= Q\ket{Q}_n, \quad Q \in \mathds{Z},\\ 
     \hat{\phi}_n\ket{Q}_n &= \ket{Q-1}_n.
     \end{align}
 \end{subequations}
 The operators $\hat{U}_n$ and $\hat{E}_{n}$ (called electric field operator) are associated with the gauge field on link $n$, which joins lattice sites $n$ and $n+1$. These operators satisfy the commutation relation~\cite{Fradkin1979,Jones1979,Gonzalez-Cuadra2017}
 \begin{equation}
     [\hat{E}_n, \hat{U}_{n'}^\dagger] = \delta_{n,n'} \hat{U}_n^\dagger,
     \label{eq:electric_field_commutation}
 \end{equation}
 and $\hat{U}_n$ is a descending operator for the eigenstates of the electric field $\hat{E}_n$:
 \begin{subequations}
     \begin{align}
     \hat{E}_n\ket{E}_n &= E\ket{E}_n, \quad E \in \mathds{Z}, \label{eq:eigenstates_E}\\
     \hat{U}_n\ket{E}_n &= \ket{E-1}_n.
     \end{align}
 \end{subequations}
 The U(1) Higgs Hamiltonian for one spatial dimension is given by~\cite{Gonzalez-Cuadra2017}
 \begin{equation}
 \begin{split}
 \label{eq:higgs_1d}
     \hat{H} &= \frac{1}{2R^2}\sum_{n=1}^N \hat{Q}_{n}^2
     -\frac{R^2}{2}\sum_{n=1}^{N-1} \left(\hat{\phi}_{n}^{\dagger} 
     \hat{U}_{n}^{\dagger} \hat{\phi}_{n+1} + \text{\text{\text{H.C.}}}\right)\\
     &+ \frac{1}{2\beta}\sum_{n=1}^{N-1} \left(\varepsilon_0 + \hat{E}_{n}\right)^2 - \beta(N-1),
 \end{split}
 \end{equation}
 where $N$ is the number of lattice sites, $\beta = 1/g^2$, $g$ is the coupling strength, and $R^2$ is inversely proportional to the mass of the Higgs field. Additionally, $\varepsilon_0$ is the background electric field. In one spatial dimension, the topological term is proportional to the background electric field, with the $\theta$ given by $2\pi\ez$.~\cite{coleman1979}. In  Eq.~\eqref{eq:higgs_1d} we have fixed the lattice spacing to be $a=1$ and throughout the paper we use natural units $\hbar = c = 1$. The first and third term in the Hamiltonian describe the Higgs field energy and electric field energy respectively. The second term is referred to as the kinetic term; it allows charge to be transferred between adjacent lattice sites, at the expense of changing the electric field between the sites. The presence of the gauge field operator in this term ensures the local gauge symmetry of the model is conserved.
 
 In the Hamiltonian formulation of the model, physical states (i.e. gauge-invariant states) have to obey Gauss's law
 \begin{equation}
 \label{eq:gauss_law}
     \hat{G}_n \ket{\Psi_{\text{physical}}} = G_n \ket{\Psi_{\text{physical}}},
 \end{equation}
 where we have defined for each lattice site $n$
 \begin{equation}
 \label{eq:gauss_law_1}
 \hat{G}_n = \hat{E}_n - \hat{E}_{n-1} - \hat{Q}_n.
 \end{equation}
 Note that the $\hat{G}_n$ operators commute with the Hamiltonian and are the generators of time-independent gauge transformations. The eigenvalues $G_n$ take integer values, and can be interpreted as static charges that can be introduced at every lattice site. For the rest of this paper we focus on the sector of vanishing static charges, $G_n=0$ $\forall n$.
 
 The Hamiltonian in Eq.~\eqref{eq:higgs_1d} conserves the total charge
 \begin{equation}
     \label{eq:q_tot}
     \hat{Q}_{\text{total}} = \sum_{n=1}^N \hat{Q}_n.
 \end{equation}
 As a result, the Hilbert space can be divided into subsectors, each containing states with a definite total charge. For the remainder of the paper, we focus on the ground-state properties of the theory in the $Q_{\text{total}} = 0$ subsector. 
 In order to ensure that the ground state is in the correct subsector during numerical calculations, we can add a penalty term to the Hamiltonian
 \begin{equation}
     \hat{H}_{\text{penalty}} = \ell\left(\sum_{n=1}^N \hat{Q}_n\right)^2. \label{eq:penalty}
 \end{equation}
 When the weight $\ell$ of this term is large enough, the states outside of the $Q_{\text{total}} = 0$ sector are penalized and removed from the low-lying spectrum. Note that while this term is necessary for matrix product state results (discussed in Sec.~\ref{sec:mps}), it is not needed for the cost function of the VQE, since we choose an initial state in the correct sector, and a variational circuit that preserves the total charge.
 
 In one spatial dimension and with open boundary conditions, it is possible to integrate out the gauge field's degrees of freedom and work with an effective Hamiltonian described by the Higgs degrees of freedom only~\cite{Hamer1997, Banuls2013, martinez_real-time_2016, muschik_u1_2017}. This procedure ensures that the eigenstates of the effective Hamiltonian are gauge invariant, and therefore allows for a contained quantum simulation, that takes place only in the gauge invariant (physical) subspace. 
 
 As described in more detail in App.~\ref{app:elimination}, the effective Hamiltonian is given by
  \begin{equation}
 \begin{split}
 \label{eq:higgs_e}
 \hat{H} &= \frac{1}{2R^2}\sum_{n=1}^N \hat{Q}_{n}^2
 -\frac{R^2}{2}\sum_{n=1}^{N-1} \left(\hat{\phi}_{n}^{\dagger} 
 \hat{\phi}_{n+1} + \text{\text{H.C.}}\right)\\
 &+ \frac{1}{2\beta}\sum_{n=1}^{N-1}(N-n)\hat{Q}_n^2
 + \frac{1}{\beta}\sum_{n=2}^{N-1}\sum_{j=1}^{n-1}(N-n) \hat{Q}_j\hat{Q}_n \\
 &+ \frac{\varepsilon_0}{\beta}\sum_{n=1}^{N-1}(N-n)\hat{Q}_{n} - \beta(N-1) + \frac{\varepsilon_0^2}{2\beta}(N-1). 
 \end{split}
 \end{equation}
 Notably, the elimination of the gauge fields introduces long-range interactions between the Higgs fields. It also allows one to simulate the model using fewer modes, as the redundant degrees of freedom have been removed, making it more accessible to quantum hardware.
 
 In order to study the phase space of the model we consider as order parameter the electric field density (EFD) of the model, which is defined as
 \begin{equation}
\begin{split}
F &= \frac{1}{N-1}\sum_{n=1}^{N-1}\expec{\hat{E}_n + \ez} \\
&= \frac{1}{N-1}\sum_{n=1}^{N-1}(N-n)\expec{\hat{Q}_n} + \ez,
\label{eq:efd}
\end{split}
 \end{equation}
 where the expectation value is taken w.r.t. the ground state.

 \subsection{Microwave-photon cavity \label{sec:platform}}
  
\begin{figure}[ht]
     \centering
     {
         \fontsize{9pt}{11pt}\selectfont%
         \def\svgwidth{\columnwidth}
         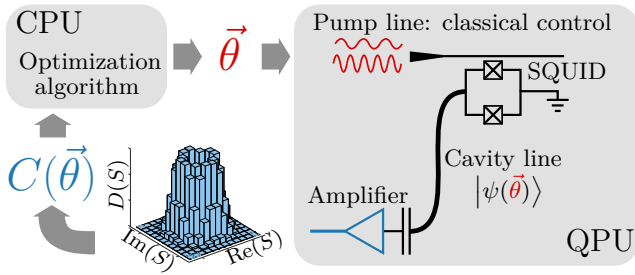
     }
     \caption{Schematics of the VQE. The quantum processing unit (QPU) consists of a microwave cavity terminated by a SQUID. The variational parameters $\vec{\theta}$ determine how the SQUID is pumped and consequently which interactions are generated between the cavity modes to generate the ansatz state $\ket{\psi(\vec{\theta})}$. A quantum-limited amplifier allows for performing heterodyne measurements on the cavity modes to obtain the cost function $C(\vec{\theta})$. In our case the cost function is the expectation value of the Hamiltonian for which we want to find the ground state, as defined in Eq.~(\ref{eq:costf}). $C(\vec{\theta})$ is used by the classical processing unit (CPU) to find the optimal variational parameters that minimizes it.}
     \label{fig:experiment}
 \end{figure}

 Here we detail the microwave-photonic hardware that we consider for our VQE of the U(1) Higgs Hamiltonian with a topological term. As shown schematically in Fig.~\ref{fig:experiment}, the quantum processor of our VQE scheme consists of a multimode coplanar waveguide resonator terminated by a SQUID at one end~\cite{chang_generating_2018, chang_observation_2020}. The SQUID is coupled both to a microwave pump mode for classical control and to the total flux of the cavity. The SQUID - consisting of a superconducting loop with two Josephson junctions - provides a high degree of dissipationless, nonlinearity in the system. This basic element is the key to the success of superconducting computing architectures~\cite{wendin2017quantum,  krantz2019quantum, blais2020circuit, vrajitoarea_quantum_2020} and is used here to control microwave fields. 
 
 The several resonant modes available in the cavity are each mapped to one of the Higgs fields as described in Sec.~\ref{sec:HOBM}, so that the original spatial lattice of the model is simulated on a lattice in synthetic dimensions. Using microwave photons as quantum degrees of freedom in a VQE has the advantage that a larger Hilbert space can be accessed compared to traditional, qubit-based VQE protocols. This property is a naturally good fit for the simulation of gauge or Higgs fields, which are defined on infinite-dimensional Hilbert spaces. As a result, using bosonic degrees of freedom requires fewer experimental modes, and avoids the translation of two-body terms into three- and four-body interactions, which usually occurs when when encoding multilevel gauge fields using qubits~\cite{Paulson:2020zjd}.  
 
 As explained in Sec.~\ref{subsec:resource_ham} and shown in Refs.~\cite{chang_generating_2018, chang_observation_2020, sandbo_thesis,hung_quantum_2021}, the SQUID in the experimental setup is an important source for nonlinear interactions, with its cosine potential providing access to a range of high-order nonlinearities. The pump mode provides a high degree of classical control over which of the available interactions are activated at a given time. Furthermore, the flexibility of circuit fabrication, allows for the design of other types of Josephson circuits, related to the SQUID, that can enhance or suppress certain nonlinearities~\cite{frattini20173}. This means one can create highly customizable gates by controlling the interacting modes, the type of interaction, its strength and relative phases. In App.~\ref{app:4op_der}, we explain in detail how to use the nonlinear operations to create interactions involving the product of four photonic operators, which are employed in our simulation of the Higgs model, but in principle these ideas can be extended much further. This flexibility is particularly suited for special purpose VQEs, including the one considered in this paper. In addition, it is natural for this platform to generate effective couplings that are nontrivial (see Sec.~\ref{subsec:resource_ham} and App.~\ref{app:4op_der} for details), which can be beneficial to other VQEs of lattice gauge theories. Overall, our approach is complementary to existing VQE schemes in the sense that it has strikingly different features than qubit-based protocols, but also uses resource Hamiltonians that are different from other bosonic platforms, such as optical photons~\cite{knill2001scheme, kok2007linear} or ultracold atoms~\cite{bloch2012quantum}.

 For a microwave cavity the density of available modes can be increased by increasing the physical length of the cavity. Within the 8GHz of measurement bandwidth, we can reach approximately 10 modes with a frequency spacing of 100MHz within 1GHz, while keeping the same order for the strength of the necessary interactions. In order to increase further the number of available modes, more microwave cavities can be coupled parametrically to each other, as has been already demonstrated with hundreds of cavities in Ref.~\cite{houck-chip_2012}. We consider this a viable path to scale the proposed system to simulate larger lattices.

 \subsection{HOBM mapping\label{sec:HOBM}}
 
\begin{figure}[ht]
     \centering
     \includegraphics[width = 0.6\columnwidth]{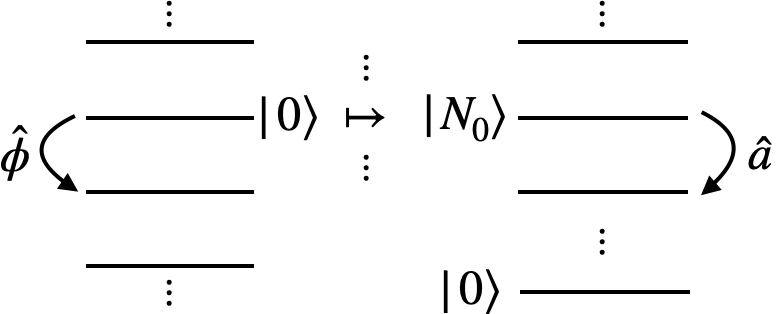}
     \caption{Schematic of HOBM mapping. The charge basis state (left) $\ket{Q}$ with charge $Q$ is mapped to the Fock state (right) with $Q+N_0$ photons $\ket{Q+N_0}$ under the HOBM mapping. In the HOBM, $\hat{a}$ is proportional to the Higgs field operator $\hat{\phi}$ according to Eq.~\eqref{eq:aphi}.}
     \label{fig:ladder}
 \end{figure}

 In order to simulate the Higgs Hamiltonian in Eq.~\eqref{eq:higgs_e} using a microwave cavity, we map its operators $\hat{Q}_n$ and $\hat{\phi}_n$ to photonic degrees of freedom using the Highly Occupied Boson Model (HOBM)~\cite{zohar_confinement_2011,Yang2016,ott_scalable_2021}, which is given by
 \begin{subequations}
     \begin{align}
     &\hat{Q}_n \rightarrow \hat{N}_n - N_0,\\
     &\hat{\phi}_n \rightarrow \frac{1}{\sqrt{N_0}}\hat{a}_n.
     \label{eq:aphi}
     \end{align}
 \end{subequations}
 In the expression above, $N_0 \in \mathds{N}$ is a constant such that the zero charge state in the Higgs model maps to the Fock state with $N_0$ photons (see Fig.~\ref{fig:ladder}). $\hat{N}$ and $\hat{a}$ are the photonic number and lowering operators, respectively. With this mapping, the U(1) Higgs Hamiltonian becomes 
 \begin{widetext}
     \begin{equation}
     \begin{split}
     \hat{H}_\text{HOBM} = &\sum_{n=1}^N\left(\frac{1}{2R^2} + \frac{N-n}{2\beta} \right)\hat{N}_n^2
     - \frac{R^2}{2N_0}\sum_{n=1}^{N-1}\left(\hat{a}_n^\dagger\hat{a}_{n+1} + \text{\text{H.C.}}\right) 
     + \sum_{n=2}^{N-1}\sum_{j=1}^{n-1}\frac{N-n}{\beta} \hat{N}_j\hat{N}_n \\
     &+ \sum_{n=1}^{N}\left(\frac{(N-n)(2\ez-N_0(N+n-1))}{2\beta} - \frac{N_0}{R^2} \right)\hat{N}_n 	   \\
     &- \beta(N-1) + \frac{\ez^2}{2\beta}(N-1) + \frac{N_0^2N}{2R^2} + \frac{N(N-1)N_0((2N-1)N_0-6\ez)}{12\beta}.
     \label{eq:Ham_HOBM}
     \end{split}	          
     \end{equation}
 \end{widetext}
The Hamiltonian given in Eq.~\eqref{eq:Ham_HOBM} is now expressed in terms of the microwave modes available in the superconducting cavity. Note that the conservation of the total charge given in Eq.~\eqref{eq:q_tot} corresponds to the conservation of the total number of photons $\sum_{n = 1}^{N} \hat{N}_{n}$.

\begin{equation}
	 F = \frac{1}{N-1}\sum_{n=1}^{N-1}(N-n)\expec{\hat{Q}_n} + \ez.
\end{equation}
 
 \subsection{Resource Hamiltonians \label{subsec:resource_ham}}
 
 In this section we discuss the resource Hamiltonians that are used in the VQE and how they can be generated by the microwave-photon platform's interaction Hamiltonian. The resources considered in the present work conserve the total number of photons, since it is a symmetry of the considered model, as discussed more in detail in Sec.~\ref{sec:HOBM}. However, we note that in general it is possible to realize also non photon number conserving interactions between photonic modes~\cite{chang_observation_2020}.
 
 By driving the pump at a frequency equal to the difference in the cavity mode frequencies of two modes $c$ and $d$~\cite{chang_observation_2020}, we can create a $c-d$ beam splitter interaction (see Ref.~\cite{sandbo_thesis} for details)
 \begin{equation}
 \hat{H}_{\text{BS}}^{(c,d)} (\vec{\Omega},\lambda) = g\left( e^{i\lambda} \hat{a}^\dagger_{c}\hat{a}_d + \text{H.C.} \right) + \sum_{n = 1}^{N} \Omega_{n} \hat{N}_n,
 \label{eq:bs_inter}
 \end{equation}
 where $g$ is the strength of the interaction, $e^{i\lambda}$ is a phase in the beam splitter interaction, and $\vec{\Omega} = (\Omega_{1}\dots,\Omega_{N})$ are the rotation frequencies of the modes obtained as the difference between the natural frequency and the frequency of a suitable rotating frame.
 This interaction induces the unitary transformation $\hat{U}(t) = e^{-it \hat{H}_{\text{BS}}^{(c,d)}}$, and we can use the time $t$, the phase $e^{-i\lambda}$ and the rotation frequencies $\vec{\Omega}$ as classical variational parameters. All three parameters can be optimized by the classical algorithm during the VQE, and while the first two can be controlled experimentally, the choice of rotation frequency can be made through a suitable change of frame of reference. Note that this interaction closely resembles the kinetic term in Eq.~\eqref{eq:Ham_HOBM}.
 
 The experimental platform considered here has proven to be able to generate higher order non linear evolutions, such as the cubic interactions demonstrated in Ref.~\cite{chang_observation_2020} and the underlying principle can be extended. In App.~\ref{app:4op_der} we show that by going into a suitable rotating frame, and by driving the SQUID at a frequency significantly lower than the other characteristic frequencies of the system, we can obtain the following effective interaction
 \begin{widetext}
     \begin{equation}
     \hat{H}_{NN} (\vec{\Omega})= g^{\prime}\left( 6\sum_{n=1}^N \hat{N}_n\left(\hat{N}_n + 2N - 1\right) + 24\sum_{n=2}^N \sum_{j=1}^{n-1} \hat{N}_j\hat{N}_n  \right) + \sum_{n = 1}^{N} \Omega_{n} \hat{N}_n.
     \label{eq:4op_inter}
     \end{equation}
 \end{widetext}
 The basic working principle to create such an interaction has been experimentally demonstrated; based on those, it is expected that this interaction can be realized with a coupling strength up to $ g^{\prime} = 10^7$~Hz~\cite{chang_observation_2020, grimm2020stabilization}. When applying the time evolution induced by the resource Hamiltonian in Eq.~\eqref{eq:4op_inter}, we use the interaction time $t$ and the frequencies $\vec{\Omega}$ as variational parameters. Note that the resource Hamiltonians $\hat{H}^{(c,d)}_{\text{BS}}$ and $\hat{H}_{NN}$ are used to prepare the variational state for the VQE as indicated in Eq.~\eqref{eq:C14}. Importantly, these resources contain terms of the form $\hat{N}_n$ and $\hat{N}_n^2$, both of which appear in the target HOBM Higgs Hamiltonian given in Eq.~\eqref{eq:Ham_HOBM}. Since $\hat{H}_\text{HOBM}$ lies in the Lie algebra spanned by the resource Hamiltonians, the VQE is able to reach the ground state of the theory~\cite{chow1940, huang1983controllability}, without the need to exactly realize it.

 \subsection{Measurements of microwave photons \label{subsec:measurements}}

 In contrast to the optical domain, a natural measurement for microwave photons is linear amplification. Here we consider, in particular, a well-established measurement available for microwave photons: the linear phase-insensitive amplifier, which simultaneously amplifies both quadratures of the input field~\cite{gambetta2017building, sandbo_thesis}. All the cavity modes are coupled out to the measurement line through a single coupling element that can be modulated~\cite{PhysRevX.10.021060} to adjust the coupling strength of all modes simultaneously. The resonator fields are then absorbed by the amplifier, as part of the measurement process. Quantum mechanics dictates that amplifying the field in a phase-insensitive manner adds at least one unit of vacuum noise to the signal, once the signal is amplified. It is therefore natural to simultaneously measure two conjugate quadratures, since this does not affect the signal-to-noise ratio. The signal obtained is equivalent to a heterodyne measurement of the cavity modes. The recent development of ultra broadband quantum-limited amplifiers (JPAs and TWPAs) improved significantly the measurement speed and signal-to-noise ratio~\cite{esposito_perspective_2021} Additionally, higher-order coherence functions can also be calculated~\cite{da_silva_schemes_2010} to obtain statistics of nonlinear measurements that cannot be performed directly (for example, photon-number measurements).
 
 The quantum state emitted from the microwave cavity is amplified in the large-gain limit (see Ref.~\cite{chang_observation_2020} for details). Measuring the field quadratures at the two output ports corresponds to simultaneously measuring the self-adjoint part $\hat{X}_S = (\hat{S} + \hat{S}^\dagger)/2$ and the anti self-adjoint part $\hat{P}_S = (\hat{S} - \hat{S}^\dagger)/(2i)$ of the signal operator
 \begin{equation}
 \hat{S} = \hat{a} + \hat{h}^\dagger.
 \end{equation}
 Here $\hat{a}$ refers to the cavity output state $\ket{\psi(\vec{\theta})}$ to be measured (see Fig.~\ref{fig:experiment}) and $\hat{h}$ is the amplifier noise mode. Note that $\hat{S}$ is a normal operator, meaning $[\hat{S}, \hat{S}^\dagger] = [\hat{X}_S, \hat{P}_S] = 0$. Furthermore, the eigenvalues of $\hat{X}_S$ and $\hat{P}_S$ correspond to the real and imaginary parts of the eigenvalues of $\hat{S}$, respectively. Repeated preparation and detection of the output state of the cavity yields a measurement distribution $D(S)$ (shown in Fig.~\ref{fig:experiment}), from which any statistical moment of $\hat{S}$ and $\hat{S}^\dagger$ can be calculated~\cite{da_silva_schemes_2010}.
 
 During the VQE of the HOBM Higgs Hamiltonian $\hat{H}_\text{HOBM}$ in Eq.~\eqref{eq:Ham_HOBM}, the cost function $C(\vec{\theta}) = \bra{\psi(\vec{\theta})}\hat{H}_\text{HOBM}\ket{\psi(\vec{\theta})}$ is calculated from the data comprising the measurement histogram $D(S)$ as follows. First, we express the expectation value of the HOBM Higgs Hamiltonian in terms of the statistical moments of the measured signal operator $\hat{S}$,
 \begin{equation}
 \begin{split}
 C(\vec{\theta}) &= \bra{\psi(\vec{\theta})}\hat{H}_\text{HOBM}\ket{\psi(\vec{\theta})} \\
 &= \sum_{n} c_{n}\expec{\hat{S}_n^\dagger\hat{S}_n} + \sum_n c'_n\expec{(\hat{S}_n^\dagger)^2\hat{S}_n^2} \\
 &+  \sum_{n,m} c''_{n,m} \expec{\hat{S}_n^\dagger \hat{S}_n\hat{S}_m^\dagger \hat{S}_m} \\
 &+ \sum_n c'''_n\expec{\hat{S}_n^\dagger\hat{S}_{n+1} + \hat{S}_n^\dagger\hat{S}_{n+1}}.
 \end{split}
  \label{eq:costf}
 \end{equation}
 The above mapping of the cost function from the target  Hamiltonian $\hat{H}_{\text{HOBM}}$ can be obtained using
 \begin{subequations}
     \begin{align}
     &\expec{(\hat{S}_n^\dagger)^j \hat{S}_n^j} = \sum_{i=0}^{j} \binom{j}{i}^2  \expec{(\hat{a}^\dagger)^j \hat{a}^j},\label{eq:calc_a2}\\
     &\expec{\hat{S}_n^\dagger\hat{S}_n \hat{S}^\dagger_m\hat{S}_m} = \expec{\hat{N}_n\hat{N}_m} + \expec{\hat{N}_n} + \expec{\hat{N}_m} + 1, \label{eq:calc_a_2da}\\
     &\expec{\hat{S}_n^\dagger\hat{S}_{n+1} + \hat{S}_n\hat{S}^\dagger_{n+1}} = \expec{\hat{a}_n^\dagger\hat{a}_{n+1} + \hat{a}_n\hat{a}_{n+1}^\dagger}, \label{eq:calc_a_2db}
     \end{align}
     \label{eq:atoS}
 \end{subequations}
 which assumes that the noise modes are in the vacuum state $\ket{0}$ (i.e. this is a quantum-limited measurement) and the noise is uncorrelated with the cavity output modes.  Finally, one can experimentally extract the value of $C(\vec{\theta})$ from the last part of Eq.~\eqref{eq:costf}. In fact, expectation values of moments of the $\hat{S}$ operators can be obtained from the measured data distributions $D(S)$ using the following expressions~\cite{eichler2012characterizing}
 \begin{subequations}
     \begin{align}
     &\expec{(\hat{S}_n^\dagger)^j \hat{S}_n^j} = \int_{S_n} |S_n|^{2j} D(S_n), \label{eq:lin_meas} \\
     &\expec{\hat{S}_n^\dagger\hat{S}_n \hat{S}^\dagger_m\hat{S}_m} = \int_{S_n,S_m} |S_n|^2 |S_m|^2 D(S_n, S_m), \label{eq:lin_meas_2da} \\
     &\expec{\hat{S}_n^\dagger\hat{S}_{n+1} + \hat{S}_n\hat{S}^\dagger_{n+1}}\nonumber \\
     &= \int_{S_n,S_{n+1}} (S_n^*S_{n+1} + S_n S_{n+1}^*) D(S_n, S_{n+1}). \label{eq:lin_meas_2db}
     \end{align}
     \label{eq:Sdistr}
 \end{subequations}
Note that the indirect measurement of $C(\vec{\theta})$ through Eq.~\eqref{eq:costf} will cause the outcome's statistical variance to be larger compared to a direct measurement of the target Hamiltonian $\hat{H}_{\text{HOBM}}$. The effect of a finite number of repeated measurement will be discussed in the next section.
 
 \subsection{Measurement budget and experimental errors \label{subsec:errors}}
 
 In a VQE, the cost function $C(\vec{\theta})$ must be computed many times by the quantum device over the course of the experiment (see App.~\ref{app:intro_vqs}). Each cost function evaluation involves the preparation of the initial state, applying a sequence of gates, and the measurement of the final VQE state (together, we refer to this as a shot). As statistical and amplification noise are present in the cost function evaluation, the number of repeated measurements of the state $M$ determines how accurately we know the cost function. In particular, for a large number of measurements the variance of the cost function is given by $\sigma^2_{H}/M$, where $\sigma^2_H$ is the intrinsic variance of the Hamiltonian for the indirect measurement obtained through Eq.~\eqref{eq:costf}. Therefore, increasing the number of measurements reduces the statistical noise of the cost function, which means the number of measurements that can be performed and the repetition rate are crucial quantities when designing an experiment. 
 
 In the following, we discuss the measurement budget for the considered microwave photon setup in more detail, as it is very different from that of common VQE platforms. The time required to prepare the initial state can be very short, for example the fastest time to prepare the Fock state $\ket{2222}$ is estimated to be on the order of $10^{-8}$s~\cite{PhysRevLett.116.020501}. The upper limit on the time required to apply all the gates in the circuit, including the beam splitter in Eq.~\eqref{eq:bs_inter} and the $\hat{H}_{NN}$ interactions in Eq.~\eqref{eq:4op_inter}, is given by the cavity decay time, which ranges between $10^{-7}$s and $10^{-4}$s~\cite{PhysRevX.10.021060}. However, some gates can be applied considerably faster than that. For a cavity with fixed coupling, the time required to measure the state is also limited by the cavity decay time, as we must allow the state to escape the cavity before measuring it. There is however the possibility to employ cavities with tunable coupling to the environment~\cite{PhysRevX.10.021060}; the cavity can be closed to have a long cavity lifetime time and opened for a fast measurement process. Assuming a tunable coupling, the readout time is negligible compared to the gate application time. 
 
 As a result, the time budget for a single shot is dominated by the coupling strength, which is on the order of $10^7$~Hz for the interactions considered  in Sec.~\ref{subsec:resource_ham}~\cite{chang_observation_2020, grimm2020stabilization}. Assuming the VQE circuit consists of about ten gates as considered in Sec.~\ref{sec:vqe}, and the time of application $t$ of the Hamiltonians given in Eqs.~(\ref{eq:bs_inter}-\ref{eq:4op_inter}) is of order $\approx\frac{1}{g}\approx \frac{1}{g^{\prime}}$, as is the case of our VQE (see Sec.~\ref{sec:vqe}), the time for a single shot is therefore on the order of $10^{-6}$s. One advantage of the microwave-photon platform over other potential VQE platforms is its ability to run continuously without the need for human intervention. As a result, ignoring the time required for classical computing in the VQE feedback loop, up to $10^{11}$ measurements can be collected in a single day. 
 
 This discussion has so far ignored the effects of experimental imperfections present on this platform. The sources of error on the microwave-photon platform are similar to other gate-based superconducting processors and as a result we can expect a similar performance~\cite{kjaergaard_superconducting_2020}. We can also expect similar improvements to the platform as large, commercial entities continue to improve and develop gate-based quantum computers. In the three steps comprising a shot - state preparation, gate application, and measurements - we therefore expect the measurement to pose the biggest concern. The main issue is that measuring higher moments of the field operators, as required in Eq.~\eqref{eq:Sdistr} for instance, requires averaging times that increase polynomially in the experimental signal-to-noise ratio~\cite{da_silva_schemes_2010}.  That is, measurements of the higher moments are increasingly sensitive to measurement noise. To understand the impact of this, we present in Sec.~\ref{sec:vqe} a realistic simulation of a VQE, which takes the measurement errors fully into account.

 \section{Microwave VQE for the U(1) Higgs model \label{sec:vqe}}

In this section we propose and classically simulate a concrete and realistic microwave VQE protocol, that allows for the study of the U(1) Higgs theory with topological term as described in Sec.~\ref{subsec:Hamiltonian}. As a specific application example, we show how to experimentally detect the first-order quantum phase transition that appears in the model for values of $R^2$ below the critical value $R^2_c$. We further discuss how future microwave-based quantum simulators can study the smoothing of the phase transition if the parameter $R^2$ approaches the critical point. As discussed in the following, below the critical point, a realistic experiment would be able to find the ground state by repeating the measurement shot up to $M=10^{7}$ times for each cost function evaluation, leading to VQE computation time of up to few hours. On the other hand, closer to the critical point the noisy measurement is more of an issue and the proposed VQE would require repetitions of at least $M=10^{9}$, therefore will benefit an increased experimental repetition rate. For a detailed description of the full phase diagram of the model will be given in Sec.~\ref{sec:phase_diagram}.
 
\subsection{Simulation and noise modelling \label{subsec:vqe_results}}
 
The detection of the considered quantum phase transition is performed by using a VQE for an approximate preparation of the ground state of Eq.~\eqref{eq:Ham_HOBM}, followed by the subsequent measurement of a suitable order parameter, namely the EFD introduced in Eq.~\eqref{eq:efd}.
 
To analyse the experimental feasibility of the observation of the targeted phase transition, we carry out classical simulations of the proposed VQE protocol. 
Our simulations include the main relevant source of imperfections, namely the noise added during amplification of the microwave signal that leads to imperfect measurement of the cost function, and a finite precision with which it is known. As detailed in Sec.~\ref{subsec:measurements}, for a given variational state $\ket{\psi(\vec{\theta)}}$ prepared in the superconducting cavity, we have access to a high number of repeated measurements $M$. This allows us in our classical simulation to approximate the probability distribution of the measurement outcome as a Gaussian distribution with mean $\expval{\hat{H}_\text{HOBM}}{\psi(\vec{\theta})}$ and variance $\frac{\sigma^2_H}{M}$. The intrinsic variance of the indirect measurement $\sigma^2_H$ can be obtained without considering the redundant noise modes in our classical simulation by suitably inverting Eqs.~\eqref{eq:atoS}.
 
In the following, we consider a lattice consisting of $N=4$ lattice sites and an HOBM mapping with $N_0=2$. For an analysis of the effects of the finite lattice size $N$ and truncation effects due to the finite value of $N_0$, see Sec.~\ref{sec:mps}. To simulate the infinite-dimensional photonic Hilbert space realized in the experiment in our simulation, we truncate the local Hilbert space of each microwave photon mode to a five-dimensional subspace. For more details on the truncation effects of our numerical description of the proposed experiment, see App.~\ref{app:truncation} (note that the values of $N$ and $N_0$ affect the capability of the experiment to observe the phase transition, while the truncated numerical description of the photon modes by a five-dimensional subspace only affects our ability to predict the outcome of the experiment).
 
\subsection{VQE detection of the phase transition\label{subsec:03}}
With $N_0=2$, the vacuum state in the model corresponds to the product state with two microwave photons in each mode $\ket{2222}$, which we choose as the initial state for the VQE. As explained in Sec.~\ref{subsec:phase_with_topological_term} and in App.~\ref{app:spike_der}, the main contributions to the ground states in the two limiting cases corresponding to small and larger values of the background field $\ez$ can be determined to be $\ket{2222}$ and $\ket{1223}$ respectively. These extremal cases motivate the choice of the variational quantum circuit:
\begin{equation}
	\hat{C}_{(1,4)} (\vec{\Omega},\vec{\theta})= \prod_{j=1}^{N_l} e^{- i \theta_{2j-1} \hat{H}_{NN}(\vec{\Omega})} e^{-i\theta_{2j} \hat{H}_{\text{BS}}^{(1,4)}(\vec{\Omega})},
	\label{eq:C14}
\end{equation}
where $N_l$ layers are applied that each involve the beam splitter interaction and the $NN$-interaction, introduced in Sec.~\ref{subsec:resource_ham}. Optimizing over $\vec{\theta}$ and $\vec{\Omega}$, the VQE protocol involves $2N_l + 4$ variational parameters (the relative beam splitter phase $\lambda$ in Eq.~\eqref{eq:bs_inter} provides a possible additional parameter, which is not used and instead is zero in this example).

\begin{figure}[th]
 	\centering
 	\includegraphics[width=0.9\columnwidth]{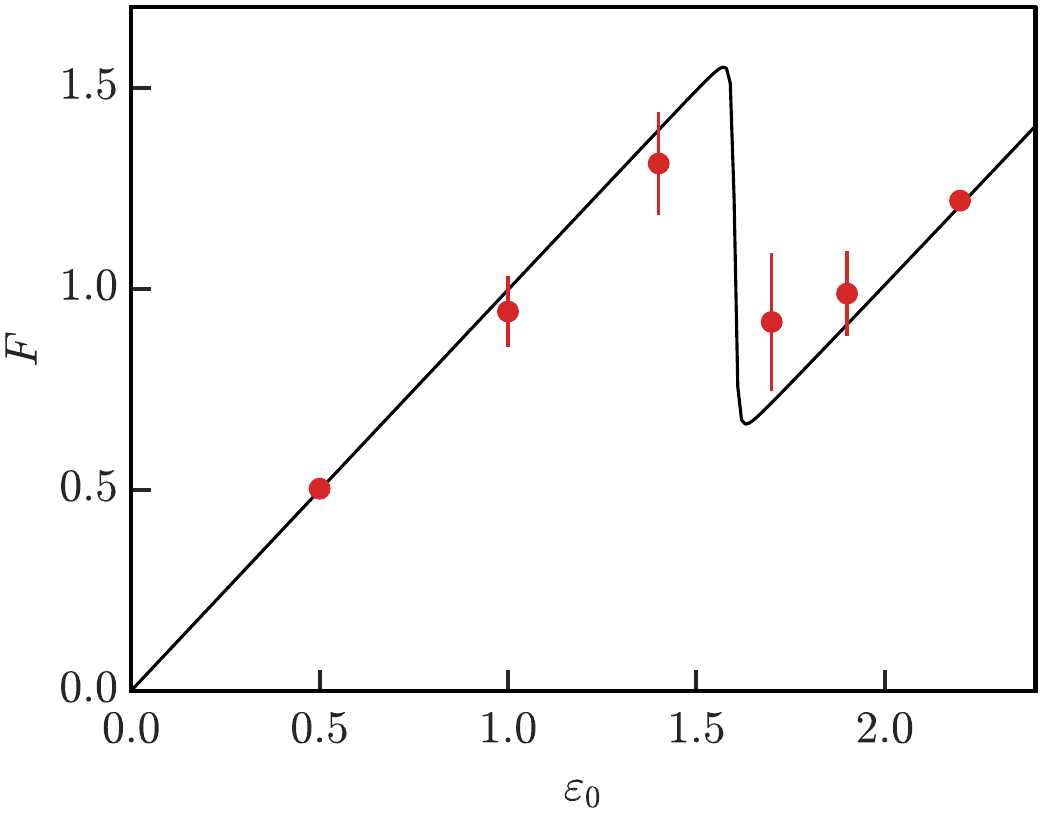}
 	\caption{EFD of the ground state for different values of $\ez$, $ R^{2}= 0.3$ and $\beta = 1$. The ground state is calculated with a classical simulation of a VQE, including the statistical noise of the cost function, assuming $M =10^{3}$ ($M = 10^{5}$) repeated measurements for points before (after) the discontinuity at around $\ez\approx1.6$. The discontinuity is a sign of the first-order transition in the ground state of the model discussed more in detail in Sec.~\ref{subsec:phase_with_topological_term}. Points and error bars are respectively the mean and standard deviation obtained with a sample of ten successful optimizations. Points after the discontinuity are post-selected as discussed in the main text. The continuous line is obtained from exact diagonalization results.}
 	\label{fig:Rsq03}
 \end{figure}

Figure~\ref{fig:Rsq03} shows that the first-order phase transition that the model undergoes for the value $R^2=0.3 < R_c^2$ can be studied in a microwave VQE experiment by observing a discontinuity in the order parameter $F$ as a function of the background field $\ez$ [$F$ is here defined as the ground state expectation value of the EFD of the model given by Eq.~\eqref{eq:efd}]. The optimization required for each point is on the order of $10^3$ cost function evaluations. We found that before (after) the discontinuity at $\ez\approx 1.6$, $M=10^3$ ($M=10^5$) measurements per cost function evaluation were required for the used approach. 

\begin{figure}[ht]
	\centering
	\includegraphics[width=0.9\columnwidth]{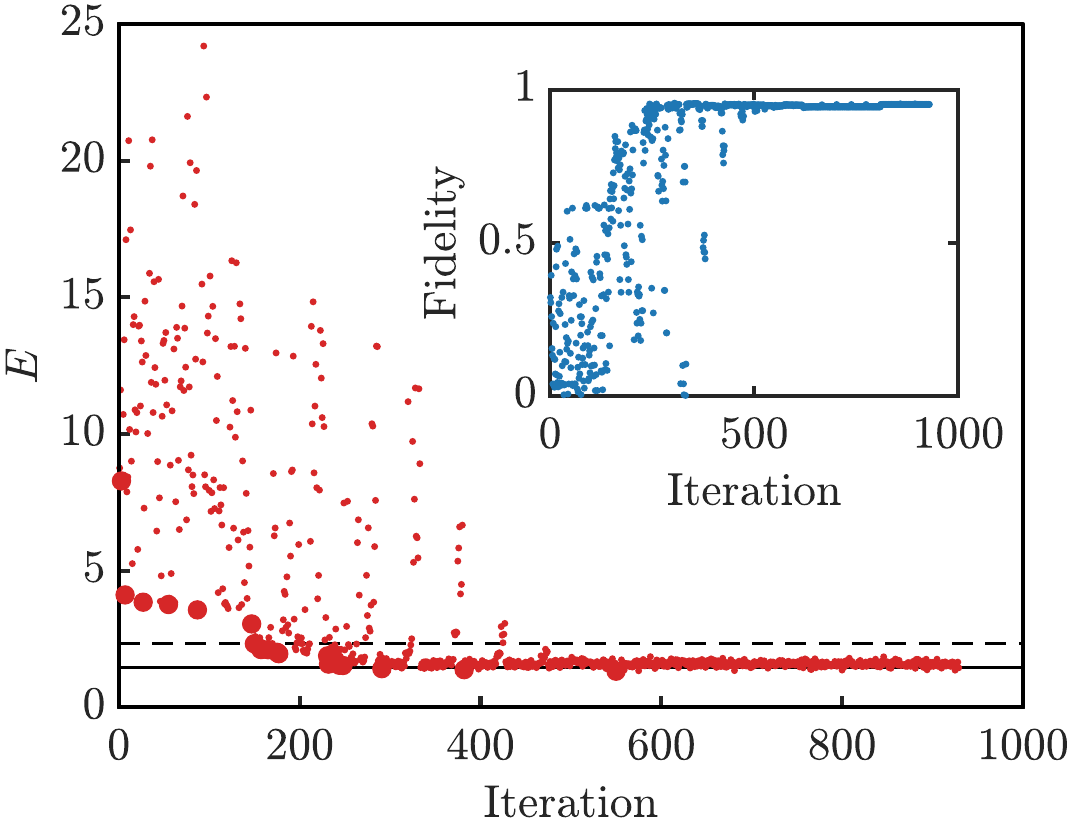}
	\caption{Cost function (expectation value of the Hamiltonian) evaluations in an successful optimization for $\ez = 1.9$. As a guide for the eye, minimum values of the cost function are represented by bigger dots. The solid (dashed) horizontal line is the ground state (first excited state) energy. Inset: fidelity of the variational state with the ground state obtained from exact diagonalization.}
	\label{fig:optimization}
\end{figure}

Points after the discontinuity are more challenging. Since we adopt a stochastic optimization algorithm, for each value of $\ez$ the optimization is repeated fifteen times, and we post-select the successful optimization as the one that has reached the lowest energy. Experimentally, this corresponds to an order of necessary shots of $10^{10}$, which is well within the feasibility limit identified in Sec.~\ref{subsec:errors}. Fig.~\ref{fig:optimization} shows an example of a successful post-selected optimization for $\ez = 1.9$. As optimizer we employ the Bayesian Adaptive Direct Search \cite{Acerbi2017Practical}, which combines a mesh-based search strategy \cite{Audet2006Mesh} with local Bayesian optimization steps \cite{Frazier2018A-Tutorial}.
 
\subsection{ VQE close to the critical point}

In the following, we explore the possibility to probe the quantum phase transition of the model in the vicinity of the critical point $R^{2} \approx R^{2}_c$, where the quantum simulation becomes most challenging. In this parameter regime, the ground state becomes highly entangled and hard to reach. For $R^2 >R_c^2$, the first-order quantum phase transition disappears, as explained in Sec.~\ref{sec:phase_diagram}.

\begin{figure}
	\centering
	\includegraphics[width=0.9\columnwidth]{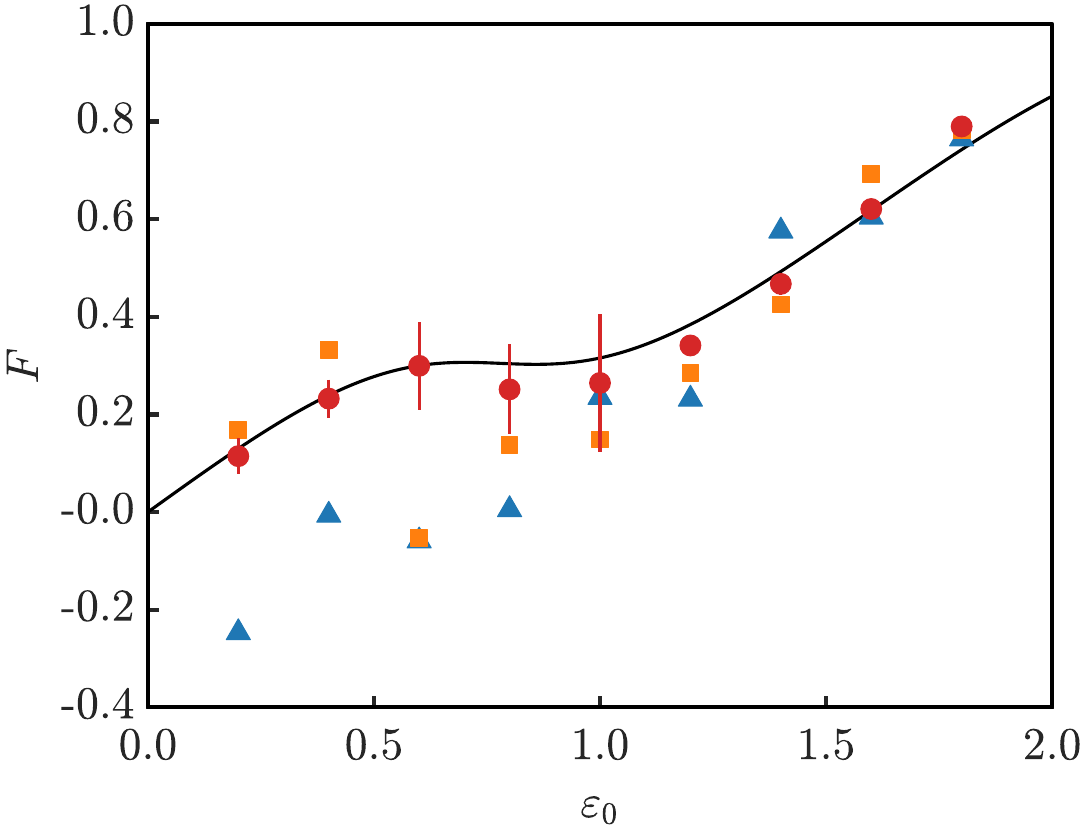}
	\caption{EFD of the ground state for different values of $\ez$, $R^{2} = 1$ and $\beta = 1$. The ground state is calculated with a classical simulation of a VQE. In this parameter regime the quantum phase transition disappears, as discussed in more detail in Sec.~\ref{subsec:phase_with_topological_term}. Data points correspond to VQE simulations with increasing number of repeated measurements for one cost function evaluation. Blue triangles: $M = 10^{6}$, orange squares: $M = 10^{9}$, red circles: $M \to \infty$ (i.e. statistical errors are neglected). }
	\label{fig:Rsq1}
\end{figure}

This behavior can in principle be probed by a VQE protocol. To illustrate this, we provide in App.~\ref{app:general_vqe} a quantum circuit that is experimentally more demanding than the example shown in Sec.~\ref{subsec:03}, but allows for the VQE preparation of the ground state of the model for the difficult parameter area around $R^2=1$. In Fig.~\ref{fig:Rsq1}, we show the classically simulated VQE result for this case including measurement noise only, i.e. in the absence of statistical errors (corresponding to $M \to \infty$). We also show how the VQE simulation with finite measurement budget approaches this limit for increasing values of the number of measurements per cost function evaluation $M=10^6$ and $M=10^9$. The plot shows that while the proposed strategy can in principle obtain the ground state in the most challenging areas of the phase diagram, noise of the cost function has a more detrimental effect than in the case of the clear first-order phase transition. For an experimental study of the ground state around $R^2=1$, it would therefore be necessary to devise a suitably adjusted measurement scheme or to increase the experimental repetition rate.

We note that in the parameter regime around $R^{2} = 1$ , the sharp discontinuity that is visible in Fig.~\ref{fig:Rsq03} is expected to become smooth. The analysis in Sec.~\ref{sec:mps} confirms that the disappearance in the discontinuity is indeed a genuine feature of the phase diagram of the model and not an artifact of finite-size or truncation effects (see Fig.~\ref{fig:mps_results_energy_field_vs_bgfield}).

  \section{Phase structure and ground-state properties\label{sec:phase_diagram}}
 In the following, we provide more information on the phase transition that is considered as concrete application example for our VQE approach in Sec.~\ref{sec:vqe}. Despite its simplicity, the U(1) Higgs model shows a rich phase diagram, especially in the presence of a topological term. For the reader's convenience in this section we briefly review the phase structure of the model, before systematically exploring how limitations of current small-scale quantum hardware affect this picture in Sec.~\ref{sec:mps}. We first focus on the case without a topological $\theta$-term in Sec.~\ref{subsec:phase_diagram_no_theta} before moving on to the phase structure in the presence of a $\theta$-term in Sec.~\ref{subsec:phase_with_topological_term}. In particular, we discuss the signature of detecting a quantum phase transition when $\ez$ in Eq.~\eqref{eq:higgs_1d} is varied and which can be observed in a quantum simulation even for small system sizes. This quantum phase transition is of particular interest for the microwave-based VQE experiments proposed in the previous section, as it is inaccessible to conventional MCMC methods due to a sign problem.

 \subsection{Phase structure in the absence of a topological term\label{subsec:phase_diagram_no_theta}}
\begin{figure}[htp!]
     \centering
     \includegraphics[width=0.98\columnwidth]{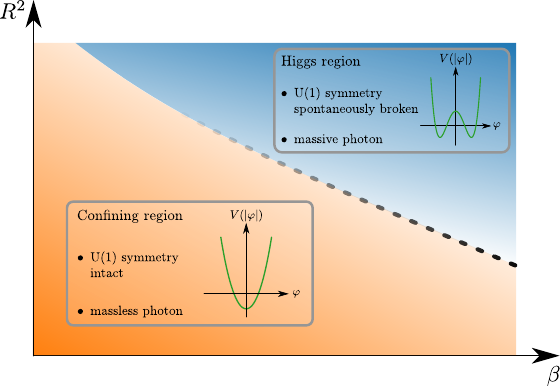}
     \caption{Phase diagram of the Abelian Higgs model with fixed length of the field in the $R^2$ -- $\beta$ plane. The Higgs region is indicated in blue whereas the confinement region is indicated in orange. The dashed line corresponds to a crossover between the Higgs and the confining region. The insets show a sketch of the form of the potential in the different regions.}
     \label{fig:pd_higgs_no_theta}    
 \end{figure}

 In the absence of a topological $\theta$-term, the Higgs model with fixed length of the field can be studied numerically with conventional MCMC lattice methods~\cite{Jones1979,Fradkin1979,Heitger1997}. An intuition for the phase structure can also be obtained by examining the Lagrangian using a simple semiclassical approach (see App. ~\ref{app:higgs_mechanism} for details). In this picture, the scalar field $\varphi$ is assumed to fluctuate only slightly around the minimum of the potential term $V(|\varphi|)$ in the Lagrangian for the ground state of the model. Depending on the value of the mass and the coupling, the shape of $V(|\varphi|)$, and consequently the nature of the ground state, changes as shown in Fig.~\ref{fig:pd_higgs_no_theta}. 
 For large values of the mass and small inverse coupling, or equivalently small values of $R^2$ and $\beta$, the potential has a unique minimum. In this region, the Abelian Higgs model essentially corresponds to a pure U(1) gauge theory, describing nothing but a massless photon. Since pure gauge theories in their compact version show charge confinement, we refer to this part of phase diagram as the confining region.
 
 Going into the opposite corner of the phase diagram in Fig.~\ref{fig:pd_higgs_no_theta}, characterized by small values of the mass and large inverse coupling, the potential $V(|\varphi|)$ shows a ``Mexican hat'' structure with an infinite number of minima along the well of the Mexican hat. In this region, the photon acquires a mass as a result of the Brout-Englers-Higgs mechanism (see App.~\ref{app:higgs_mechanism} for details) and we refer to this part as the Higgs region. The ground state in the Higgs region corresponds to one of the minima of the Mexican hat potential and, thus, spontaneously breaks the U(1) symmetry of the model. 
 
 On a lattice with finite spacing, these two regions of the phase diagram are separated by a crossover that is disappearing for small values of the inverse coupling $\beta$. Taking the continuum limit, the crossover line ends in a Berezinskii-Kosterlitz-Thouless transition from a confining phase~\footnote{Strictly speaking, one should refer to ``regions'' in the phase diagram, since there is an analytical connection between the different regions in the phase diagram} for large masses to the Higgs phase at small values of the mass~\cite{Heitger1997}. 
 
 \subsection{Phase structure in the presence of a topological term\label{subsec:phase_with_topological_term}} 
\begin{figure}[htp!]
     \centering
     \includegraphics[width=0.98\columnwidth]{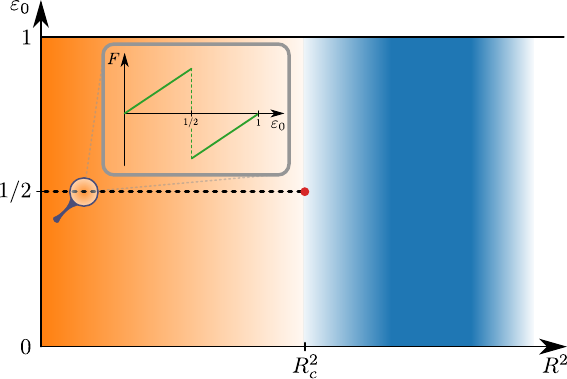}
     \caption{Sketch of the phase diagram in the $R^2$ -- $\varepsilon_0$ plane at a fixed value of $\beta$. The dashed line indicates the critical line corresponding to a first-order phase transition which ends in a second-order quantum phase transition (indicated by the red dot). The confining region is indicated in orange, the Higgs region in blue. The inset shows the behavior of the EFD which, when crossing the phase transition, is showing a jump as a result of the first-order transition (see discussion in the main text). This behavior of the EFD serves as the signature of the phase transition for the here performed quantum simulation of the model.}
     \label{fig:cylinder_phase_diagram}
 \end{figure}

 The phase structure of the model in the presence of a topological term has been investigated both theoretically~\cite{Komargodski2019} and numerically~\cite{Gattringer2018,Goschl:2018uma,sulejmanpasic2020first}. Fig.~\ref{fig:cylinder_phase_diagram} shows a sketch of the phase diagram as a function of $\varepsilon_0$ and $R^2$ for a fixed value of $\beta$. 

Most notably, the physics is periodic in $\varepsilon_0$ with period $1$. The origin of this periodicity can be understood intuitively by realizing that it is always possible to find an integer $k$, such that  $\varepsilon_0' := k-\varepsilon_0 \in[0,1)$. Inserting this into Eq.~\eqref{eq:higgs_1d} we find for the electric-energy term
\begin{align*}
    \sum_{n=1}^{N-1} \left(\hat{E}_{n} + \varepsilon_0'\right)^2 &= \sum_{n=1}^{N-1} \left(\hat{E}_{n} + k-\varepsilon_0\right)^2 \\ 
    &= \sum_{n=1}^{N-1} \left(-\hat{E}_{n}' + \varepsilon_0\right)^2
\end{align*}
while the other terms stay unchanged. The shifted electric field $\hat{E}_{n}' :=  \hat{E}_{n} + k$ fulfills the same algebra as $\hat{E}_n$ in Eq.~\eqref{eq:electric_field_commutation} and, thus, has the same spectrum as the original electric field operator (see Eq.~\eqref{eq:eigenstates_E}). In other words, $\hat{E}_{n}'$ is unitarily equivalent to $\hat{E}_n$. Moreover, since the spectrum of $\hat{E}_{n}'$  is given by the integer numbers $\mathds{Z}$, $-\hat{E}_{n}'$ is unitarily equivalent to $\hat{E}_{n}'$. As a result, $\hat{E}_{n} + \varepsilon_0'$  and $\hat{E}_{n} + \varepsilon_0$ are related by a unitary transformation, thus showing that an integer shift in the background field does not change physics and we can restrict ourselves for the following discussion to $\varepsilon_0\in[0,1)$. Moreover, the two points $\varepsilon_0 = 0,\, 1/2$ are special, because for these cases the Hamiltonian is symmetric under charge conjugation, meaning the exchange of particles and antiparticles (rigorous proofs for the periodicity and the charge conjugation symmetry are provided in App.~\ref{app:symmetries}).
 
To get further insight into the phase diagram in Fig.~\ref{fig:cylinder_phase_diagram}, we can examine the Hamiltonian in Eq.~\eqref{eq:higgs_1d} (or its equivalent version in Eq.~\eqref{eq:higgs_e}) in the limiting cases of large and vanishing mass of the Higgs field. Focusing first on the limit of small mass corresponding to $R^2\gg 1$, Eq.~\eqref{eq:higgs_e} reduces to
\begin{align}
    -\frac{R^2}{2}\sum_{n=1}^{N-1}(\hat{\phi}_n\hat{\phi}_{n+1}^\dagger + \text{H.C.}),
\end{align}
which is a pure hopping Hamiltonian. In particular, we see that the expression above is independent of $\ez$ and thus the physics does not depend on the topological term. Since we focus on the sector of vanishing total charge, the ground state in this limit is given by a superposition of all zero-total-charge states.
 
Looking at the opposite limit of large mass, or equivalently $R^2\ll 1$, the Hamiltonian in Eq.~\eqref{eq:higgs_1d} reduces to
\begin{align}
    \frac{1}{2R^2}\sum_{n=1}^N \hat{Q}_{n}^2 + \frac{1}{2\beta}\sum_{n=1}^{N-1} \left(\ez + \hat{E}_{n}\right)^2.
    \label{eq:Hamiltonian_vanishing_mass_limit}
\end{align}
The first term energetically favors a vanishing charge at every site, $Q_n=0$ $\forall n$. Inserting this into Gauss Law in Eq.~\eqref{eq:gauss_law2} and taking into account that we focus on the sector of vanishing static charges, we see that for gauge invariant states the electric field has to take a constant value for all sites. For $\ez < 1/2$ the field configuration minimizing the electric energy in Eq.~\eqref{eq:Hamiltonian_vanishing_mass_limit} is given by $E_n = 0$ $\forall n$. In contrast, for $\ez> 1/2$ an electric field of $E_n = -1$ $\forall n$ minimizes the energy. Exactly at the point $\ez = 1/2$ both configurations yield the same electric energy contribution, and the ground state is doubly degenerate. These considerations show that in the limit of large masses the model undergoes a first-order quantum phase transition as we increase $\ez$ from $0$ to $1$. This transition is accompanied by a discontinuity in the EFD (defined in Eq.~\eqref{eq:efd}) which we expect for $R^2\ll 1$ to behave as 
\begin{equation}    
    F \approx \begin{cases} \ez & \mbox{ for } \ez \leq 1/2\\
        \ez - 1 & \mbox{ for } \ez > 1/2,
    \end{cases}
\end{equation}
as shown in the inset of Fig.~\ref{fig:cylinder_phase_diagram}. Moreover, the abrupt jump in the electric field configuration for $\ez=1/2$ by one unit leads to a cusp in the electric-energy term. Hence, we expect the ground-state energy to show a non-smooth behavior at the transition point too. These expectations, which will be verified in Sec.~\ref{sec:mps}, serve as the signature of the phase transition for the microwave-based quantum simulation of the model we perform.

Our discussion of the two limiting cases shows that the behavior of the model has to change going from large masses (corresponding to small $R^2$) to small masses (corresponding to large $R^2$). While for small masses a first-order quantum phase transition occurs for $\ez = 1/2$, this transition vanishes in the limit of large mass. Ref.~\cite{Komargodski2019} argued that the critical line at $\ez = 1/2$ ends in a second-order quantum phase transition belonging to the Ising universality class at a critical value $R_c^2$. The second-order transition at the endpoint of the critical line is accompanied by a spontaneous breaking of the charge conjugation symmetry, and has been observed in numerical MCMC simulations using a dual lattice formulation of the model~\cite{Goschl:2018uma,Gattringer2018}. Moreover, despite the new features arising from the $\theta$-term, the picture from the previous section remains true for large enough values of inverse coupling. For large mass, or equivalently small $R^2$, there is a confining region whereas for large values of $R^2$ again the Higgs region occurs (see Fig.~\ref{fig:cylinder_phase_diagram}).

The microwave-based VQE approach detailed in Sec.~\ref{sec:vqe} can be used to observe the first-order phase transition of the model in the presence of a topological term, as shown in Sec.~\ref{sec:vqe}. Since current quantum hardware is of small-scale, we will restrict ourselves to a small number of degrees of freedom and we have to work with lattices consisting only of a few sites. It can be shown that even for such a small system the considerations above hold true, but the location of the first phase transition shifts from $\ez = 1/2$  to some larger value of $\ez$. Focusing on the regime well below $R^2_c$ and assuming that the mass of the Higgs field is large enough that the kinetic term can be neglected, we show in App.~\ref{app:spike_der} that the first phase transition occurs at
\begin{equation}
    \ez = \frac{1}{2} + \frac{\beta}{R^2(N-1)}. 
    \label{eq:jump_formula}
\end{equation}
Equation~\eqref{eq:jump_formula} predicts the phase transition cannot occur before $\ez = 1/2$ for a finite lattice and only in the $N \to \infty$ limit, the phase transition occurs exactly at $\ez = 1/2$. Moreover, the derivation shows that in this regime the the ground state before the phase transition is dominated by the state with vanishing charge at every site, $\ket{0}^{\otimes N}$, whereas after the transition it holds a pair of charges with opposite sign and is dominated by $\ket{-1}\ket{0}^{\otimes N-2}\ket{+1}$.

 \section{Spin Truncation and Matrix Product States\label{sec:mps}}
 
 In this section we analyse the prospects for exploring the physics of the model using existing and near-term small-scale quantum hardware. Since the number of modes available in a microwave cavity in the near future will be limited, we want to study in the following section if the limited size of the system that is studied allows to observe relevant physics. To assess the feasibility of such an approach we first explore the effects of truncating the model to a small number of degrees of freedom numerically using matrix product states (MPS). For a system with $N$ sites and open boundary conditions, the MPS ansatz reads
 \begin{equation}
    \ket{\psi} = \sum_{i_1,i_2,\dots,i_N}^d M^{i_1}_1M^{i_2}_2\cdots M^{i_N}_N\ket{i_1}\otimes\ket{i_2}\otimes\dots\otimes\ket{i_N}.
 \end{equation}
 In the expression above, $M^{i_k}_k$ are complex square matrices of size $\chi$ for $1<k<N$, and $M^{i_1}_1$ ($M^{i_N}_N$) is a $\chi$-dimensional row (column) vector. The states $\{\ket{i_k}\}_{i_k=1}^d$ are a basis for the $d$-dimensional local Hilbert space on site $k$. The parameter $\chi$ is called the bond dimension of the MPS and determines the number of variational parameters in the ansatz and limits the amount of entanglement that can be present in the state (see Refs.~\cite{Verstraete2008,Schollwoeck2011,Orus2014a} for detailed reviews). The optimal set of tensors can be found variationally by iteratively minimizing the energy for each tensor while keeping the others fixed~\cite{Verstraete2004}.
 
 For numerical calculations with MPS, the dimension of the local Hilbert spaces has to be finite, which is in contrast to the infinite-dimensional degrees of freedom for each site in the Higgs Hamiltonian. Hence, they have to be truncated to a finite dimension. In Sec.~\ref{sec:spin_trunc} we discuss a way of truncating the Hilbert spaces before numerically exploring finite-size effects on the phase structure in the presence of a $\theta$-term in Sec.~\ref{sec:mps_plots}.
 
 \subsection{Spin truncation \label{sec:spin_trunc}}
 
 One possibility of truncating the theory is to replace the bosonic fields with integer spins
 \begin{align}
     \hat{Q}_n \rightarrow \hat{S}_n^z,\quad\quad\hat{\phi}_n^\dagger \rightarrow \frac{1}{|S|}\hat{S}_n^+, 
     \label{eq:spin_map}
 \end{align}
 where $\hat{S}^z$ and $\hat{S}^+$ are the $z$-component and raising operators for a particle with spin $s$, respectively, and $|S| = \sqrt{s(s+1)}$. The resulting local Hilbert space is finite and has dimension $d=2s+1$. Using this mapping, the commutation relations of the original model stay intact (see Sec.~\ref{subsec:Hamiltonian} for more details on the commutation relations) except for
 \begin{equation}    
     [\hat{\phi}_n, \hat{\phi}_{n'}^\dagger] \to \left[\frac{1}{|S|}\hat{S}_n^-,\frac{1}{|S|}\hat{S}_{n'}^+\right] = \frac{2}{|S|^2}\delta_{n,n'}\hat{S}^z_n \label{eq:comm_spin}.
 \end{equation}
 Notice that the expression above approaches the correct commutation relation for bosonic field operators, $[\hat{\phi}_n, \hat{\phi}^\dagger_{n'}] = 0$, in the limit $s\to\infty$. The resulting spin Hamiltonian after applying the mapping in Eq.~\eqref{eq:spin_map} to Eq.~\eqref{eq:higgs_e} can be addressed with MPS using standard methods, despite its long-range interactions~\cite{Banuls2013,Banuls2016a,Sala2018,Funcke2019,Banuls2019}.
 
 \subsection{MPS results \label{sec:mps_plots}}
  
\begin{figure}[htp!]
     \centering
     \includegraphics[width=\columnwidth]{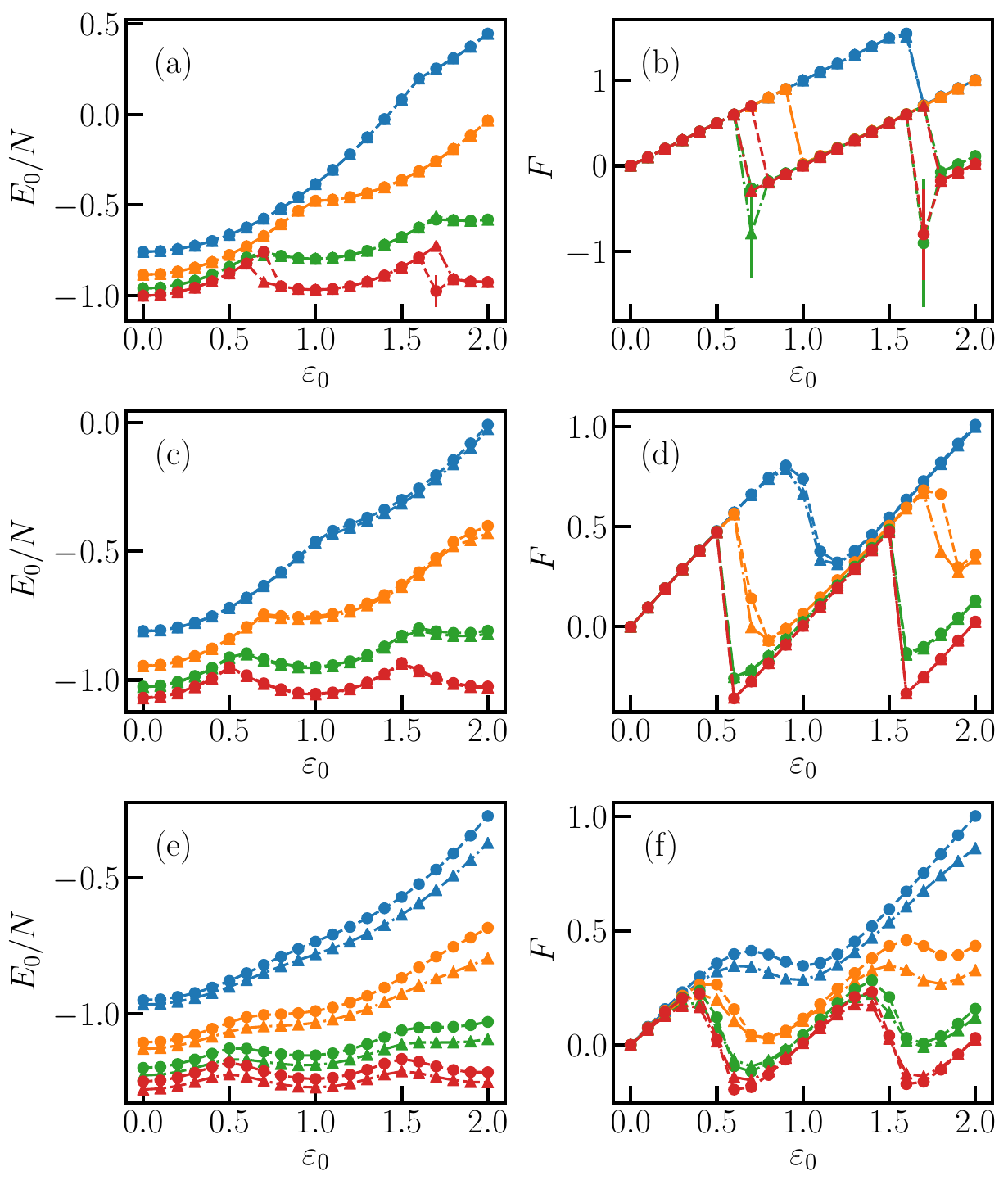}
     \caption{Ground state energy (left column) and EFD (right column) as a function of $\ez$ calculated using MPS for $\beta = 1.0$ and $R^2=0.3$ (first row), $0.6$ (second row) and $1.0$ (third row). Dots indicate $s=1$, triangles $s=2$ and the different colors encode the different system sizes $N=4$ (blue), $8$ (orange), $20$ (green) and $100$ (red). As a guide for the eye the markers are connected with lines. The error bars represent the uncertainty from the extrapolation in $\chi$.}
     \label{fig:mps_results_energy_field_vs_bgfield}
 \end{figure}

 In order to examine the effects of truncating the model to a small, finite number of degrees of freedom on the phase structure, we study the spin Hamiltonian at fixed coupling strength for a wide range of values of $R^2$, $s$, and $N$. To estimate the error due to the finite size $\chi$ of the matrices in our numerical MPS simulations, we repeat the calculation for every combination of $(R^2, N, s)$ for a range of bond dimensions $\chi\in[10;100]$. Afterwards we can extrapolate the results to the limit $\chi\to\infty$ following Ref.~\cite{Funcke2019}. Figure~\ref{fig:mps_results_energy_field_vs_bgfield} shows the MPS results obtained for the ground-state energy density $E_0/N$ and the EFD as a function of $\ez$ for various system sizes and couplings. For all the results presented, we have chosen a penalty strength of $\ell = 3N$ which is sufficient to ensure that we are in the sector of vanishing total charge. 

In general, we observe that truncation effects due to finite value of $s$ are rather small compared to finite-size effects. In particular, for larger masses (corresponding to small $R^{2}$) even the simplest nontrivial truncation, $s=1$, is sufficient to give the correct qualitative behavior. On the other hand, for the smallest masses we study, corresponding to $R^2 = 1.0$, results for $s=1$ and $s=2$ start to have a larger difference, in particular for larger values of $\varepsilon_0$, and it is where the finite truncation shows the biggest effect.

 
 In contrast to the spin truncation, finite-size effects are more pronounced. While we expect the physics to be periodic in $\ez$ with period $1$ as outlined in Sec.~\ref{subsec:phase_with_topological_term}, Fig.~\ref{fig:mps_results_energy_field_vs_bgfield} shows that ground-state energy density as well as the EFD only show perfect periodicity throughout the entire range of $\ez$ we study for our largest system size, $N=100$. For smaller values of $N$ the characteristic features still repeat, however the graphs for the energy density and the EFD show an overall increasing trend with $\ez$. Moreover, the period at which the characteristic features repeat is increasing with decreasing system size.
 
 Focusing on our results for $R^2=0.3$ in Figs.~\ref{fig:mps_results_energy_field_vs_bgfield}(a) and~\ref{fig:mps_results_energy_field_vs_bgfield}(b), corresponding to the largest value of the mass we consider, we clearly see the signatures of the first-order phase transition as discussed in Sec.~\ref{subsec:phase_with_topological_term}. The EFD shows sharp discontinuities accompanied by cusps in the ground-state energy for all system sizes we study. For our largest system size, $N=100$, these transitions occur for $\ez$ close to integer multiples of $1/2$. Decreasing the system size, we observe that the location of the first transition gradually shifts to larger values of $\ez$, in agreement with Eq.~\eqref{eq:jump_formula}. In particular, even for an extremely small system size of $N=4$ the signatures of phase transition are still clearly visible in both the EFD and the ground-state energy density.
 
 Going to a significantly smaller mass, the discontinuities and cusps in the EFD as well as the ground-state energy density vanish, as our data for $R^2=1.0$ in Figs.~\ref{fig:mps_results_energy_field_vs_bgfield}(e) and~\ref{fig:mps_results_energy_field_vs_bgfield}(f) reveal. The smooth behavior of these observables is giving a clear indication that we are in the regime below the critical mass, or correspondingly $R^2>R^2_c$, and the phase transition is gone. In general, finite-size effects seem to shift the critical mass at which the transitions vanishes towards smaller values (or equivalently $R_c^2$ towards higher values), as our data for $R^2=0.6$ reveals. 
 Looking at Figs.~\ref{fig:mps_results_energy_field_vs_bgfield}(c) and~\ref{fig:mps_results_energy_field_vs_bgfield}(d) we clearly observe a  a first-order order transition signaled by a sharp discontinuity in the EFD for system sizes $N\geq 20$. In contrast, the discontinuities are no longer present for the two smallest system sizes $N=4,8$ we study, and the EFD curves for these cases are smooth.
 
\begin{figure}[htp!]
     \centering
     \includegraphics[width=\columnwidth]{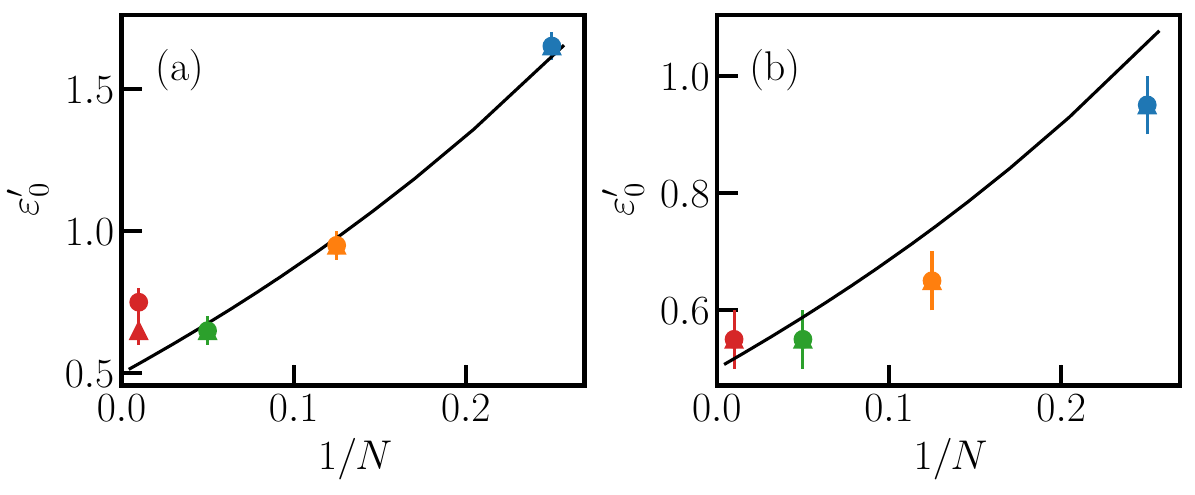}
     \caption{Location of the first discontinuity extracted from our numerical data as a function of system size for $\beta = 1.0$ and $R^2=0.3$ (a), and $0.6$ (b). Dots indicate $s=1$, triangles $s=2$, and the solid line the prediction from Eq.~\eqref{eq:jump_formula} for the large mass limit. The error bars originate from our finite resolution in $\ez$.}
     \label{fig:mps_results_jump_location_vs_size}
 \end{figure}

 From our data in Figs.~\ref{fig:mps_results_energy_field_vs_bgfield}(a) -~\ref{fig:mps_results_energy_field_vs_bgfield}(d) we can extract the location $\ez'$ of the first-order phase transition as the first discontinuity (cusp) in the EFD (energy density). Fig.~\ref{fig:mps_results_jump_location_vs_size} contains our results for $R^2 = 0.3$, $0.6$ and various system sizes.
 Both panels show that finite-size effects shift the location of the first transition to values $\ez'>1/2$. Comparing our numerical data to the prediction for the large mass limit $R^2\ll 1$ in Eq.~\eqref{eq:jump_formula}, we observe good agreement for $R^2=0.3$. Going to a larger value of $R^2=0.6$, or equivalently to a smaller mass, the approximation in Eq.~\eqref{eq:jump_formula} breaks down and is no longer compatible with the numerical data.
 
 In summary, our MPS data demonstrate that we can observe the relevant features of the model, even if we restrict ourselves to a small number of degrees of freedom. The truncation of the bosonic fields does not severely affect the physics in the presence of a $\theta$-term for the range of parameters we have considered. In particular, even for very small system sizes that are accessible with current quantum hardware, we can observe the characteristic features of the phase structure. For large masses, or equivalently small values of $R^2$, the EFD and the ground-state energy density clearly show the signatures of the first-order phase transition which eventually vanish as we go to small masses or correspondingly large $R^2$. We note that while for the parameters we have considered the MPS data is a useful benchmark for the proposed VQE, in general the VQE setup allows for going beyond these regimes, and, thus, for exploring physics that is not easily accessible with classical simulations.

 \section{Conclusions\label{sec:conclusion}}

We have designed a VQE protocol to be run on a superconducting microwave cavity. Here, the crucial advantage is the provision of bosonic degrees of freedom which allows to implement an infinite-dimensional Hilbert space. In our case, truncation of different sizes are available for quantum simulation, offering the possibility to study and control the effect of the finite truncation. For a proposal to simulate gauge fields using the qudit space available in Rydberg atoms see Ref.~\cite{gonzalez-cuadra_hardware_2022}. The readout of such a system is inherently different when compared to standard qubit-based simulations -- it allows us to calculate different statistical moments from one single collection of measurement data. Moreover, it can be run continuously without the need for periodic, time-consuming recalibrations of the control electronics. 
Here we have shown that this platform is able to experimentally demonstrate an interesting feature of the lattice Higgs model with a topological term, namely its first-order quantum phase transition. In order to determine that indeed the physics of this model is probed, we used classical simulations based on matrix product states to study the effects of a finite lattice size and truncated operators.

The application of quantum simulations to lattice gauge theories has the potential to extend its reach beyond what has been possible so far with classical numerical methods, and has been a quickly developing field \cite{klco_su2_2020,ciavarella_trailhead_2021,Paulson:2020zjd,Haase:2020kaj,atas_su2_2021,banuls_simulating_2020,zohar_quantum_2021,kasper_jaynes-cummings_2020,riechert_engineering_2022}. More specifically, studying lattice gauge theories that include topological terms is of particular interest, since they generate rich and interesting physics, and they are one of the settings that are, in general, intractable with conventional MCMC lattice methods. Our implementation of a bosonic VQE  paves the way to extend this work in order to study topological terms in higher dimensions, with the ultimate goal to reach 3+1 dimensions. A work along these lines is~\cite{kan_investigating_2021}. To that end, a VQE of a high-energy lattice gauge theory in three spatial dimensions is an important and ambitious milestone in the field of quantum simulation and a fundamental step towards answering some open questions about the universe. We note that in the case of higher number of spatial dimensions only part of the gauge fields can be eliminated through Gauss' law. This was done for example in Refs.~\cite{Paulson:2020zjd,Haase:2020kaj,kan_investigating_2021}. The remaining gauge degrees of freedom can be mapped appropriately to photonic modes similarly to what was done for the Higgs field in the present work, and there are no fundamental limitations to extending our work to higher spatial dimensions. An interesting extension of this work will be the study of real-time evolution of the lattice Higgs model in 1+1 dimensions. More specifically, the resource Hamiltonians presented in Sec.~\ref{subsec:resource_ham} are extremely close to the terms appearing in the Hamiltonian, thus potentially offering the possibility of realizing a trotterized time evolution. Interesting dynamical effects include, for example, string breaking or the study of the phase diagram of the Higgs model with quenched systems~\cite{Kuno2017}.

For near term applications, there are many exciting opportunities to use the unique features of the microwave cavity platform (as explained in Sec.~\ref{sec:platform}) in upcoming VQE experiments. For example, the type of measurements allows for calculating higher-order correlation functions without requiring a change in the measurement protocol. This is in stark contrast to qubit-based VQEs, which are readout via the measurement of different Pauli strings. Their number quickly increases as soon as higher-order correlation functions should be accessed, which usually renders the experiment infeasible due to the, then necessary, exorbitant measurement budget.
A further possibility that we have not yet exploited in the current work is the ability to perform simultaneous, two-tone pumping on the platform~\cite{chang_generating_2018}. This would allow for effectively combining the evolution generated by two resource Hamiltonians, without the need of approximating it with successive evolutions. Such technique could greatly expand the types of interactions available for future VQE implementations. 
Another promising path would be hybrid quantum platforms that include both qubits and bosons on the same hardware. In particular, these are an excellent fit for a lattice gauge theory simulation that involves fermionic matter fields. Given the highly tunable interactions available on this platform, it would also be suitable for VQEs of models outside of lattice gauge theories, including quantum chemistry and condensed matter physics.

\begin{acknowledgments}
    This work has been supported by Transformative Quantum Technologies Program (CFREF), NSERC, New frontiers in Research Fund, European Union’s Horizon 2020 research and innovation programme under the Grant Agreement No.\ 731473 (FWF QuantERA via QTFLAG I03769). C.A.M.\ acknowledges the Alfred P.\ Sloan foundation for a Sloan Research Fellowship.
    S.K.\ acknowledges financial support from the Cyprus Research and Innovation Foundation under project ``Future-proofing Scientific Applications for the Supercomputers of Tomorrow (FAST)'', contract no.\ COMPLEMENTARY/0916/0048.
    J.F.H. acknowledges support by the ERC Synergy grant HyperQ (Grant No. 856432). 
    C.M.W.\ acknowledges Canada First Research Excellence Fund (CFREF), NSERC of Canada, the Canadian Foundation for Innovation, the Ontario Ministry of Research and Innovation, and Industry Canada for financial support.
\end{acknowledgments}

 
 \appendix
 
 \section{Introduction to VQEs \label{app:intro_vqs}}
 
 In the following we review the basic principles of variational quantum eigensolver (VQE) algorithms. For more details we refer to the review in Ref.~\cite{mcclean_theory_2016} and the initial demonstrations in \cite{Peruzzo2014}. 
 In the work at hand, a VQE is employed to approximate the ground state of the target Hamiltonian $\hat{H}_T$, utilizing a closed feedback loop between the quantum processor and a classical optimizer.
 
 The algorithm may be summarized as follows. First, the quantum device is initialized with an easily preparable state $\ket{\psi_\text{in}}$. Next, a sequence of gates is applied to the initial state. The gates in this sequence are the unitaries $ \exp(-i\theta_k\hat{H}^{(j)}_R)$, where $\hat{H}^{(j)}_R$ are the available resource Hamiltonians. Note that the set of available interactions does not need to be a universal gate set; a restricted gate set specific to the target Hamiltonian or the problem at hand is sufficient. The $\theta_k$ are the variational parameters; their values are chosen by the classical optimizer and passed to the quantum device. In the circuit, the $\theta_k$ manifest as the product of the interaction strength of $\hat{H}^{(j)}_R$ and the time for which the gate is applied. In a typical VQE circuit, a certain sequence of gates is often repeated with different variational parameters, where each elementary sequence of gates that is repeated is called a layer.
 
 Once the VQE circuit has been applied to the initial state, the result is the VQE ansatz state $\ket{\Psi(\vec{\theta})}$. For example, if the circuit employs $r$ resource Hamiltonians in each layer and $L$ layers in the circuit, then 
 \begin{equation}
     \label{eq:vqs_circuit}
     \ket{\Psi(\vec{\theta})} = \prod_{\ell=1}^L\prod_{j=1}^r\exp\left(-i\theta_{j+r(\ell-1)}\hat{H}^{(j)}_R \right)\ket{\psi_\text{in}}.
 \end{equation}
By tuning the the parameters $\vec{\theta}$ to their correct (but not necessary unique) values, this state approximates the the ground state of $\hat{H}_T$, provided the ansatz is expressive enough.. 
Once the ansatz state is prepared for a given set of variational parameters, the cost function $\bra{\Psi(\vec{\theta})}\hat{H}_T\ket{\Psi(\vec{\theta})}$ is evaluated on the quantum device. Since quantum measurements are inherently stochastic, the preparation and measurement of $\ket{\Psi(\vec{\theta})}$ needs to be repeated in order to obtain a precise estimate of the cost function. In particular, if the state is measured $M\gg1$ times and $\sigma^2_H = \expec{\hat{H}_T^2}-\expec{\hat{H}_T}^2$ is the variance of the target Hamiltonian w.r.t. $\ket{\Psi(\vec{\theta})}$, then the variance in the cost-function estimate is determined by $\sigma^2_{H}/M$. 
The estimate of the cost function is fed to a classical processor which employs an optimization algorithm to choose new values for the variational parameters, that are supposed to decrease the cost. These new parameters are fed back into the quantum device, where the process begins anew. This cycle is repeated until the convergence threshold for the classical optimizer is reached or the allocated measurement budget is exhausted.
 
 \section{Elimination of the gauge fields}
 \label{app:elimination}
 
 In the following, we express the Hamiltonian in Eq.~\eqref{eq:higgs_1d} solely in terms of Higgs degrees of freedom. Generally, the complete elimination of the gauge field is always possible if only one spatial dimensions and open boundary conditions are considered. Such transformation results in an effective Hamiltonian that requires fewer modes to be simulated on the quantum hardware.
 
 By considering the subspace of the theory that corresponds to a zero static charge on each vertex, $\hat{G}_n\ket{\Psi_{\text{physical}}} = 0$, Eq.~\eqref{eq:gauss_law_1} can be rewritten as
 \begin{equation}
 \label{eq:gauss_law2}
     \hat{E}_n =\hat{E}_{n-1} +\hat{Q}_n
 \end{equation}
 for the physical states (see Sec.~\ref{subsec:Hamiltonian}).
 
 Employing open boundary conditions,  Eq.~\eqref{eq:gauss_law2} can be solved recursively to obtain the electric field at every site resulting in
 \begin{equation}
 \label{eq:gauss_sol1d}
     \hat{E}_n = \sum_{k=1}^n \hat{Q}_k.
 \end{equation}
 This expression allows us to express the electric field operators entirely in terms of the charge operators. In particular, the electric term of the Hamiltonian can be rewritten as
 \begin{equation}
 \begin{split}
 \label{eq:e_der}
     \sum_{n=1}^{N-1} \left(\varepsilon_0 + \hat{E}_n\right)^2  =& \sum_{n=1}^{N-1} \left(\varepsilon_0 + \sum_{k=1}^n \hat{Q}_k\right)^2 \\
     =& \sum_{n=1}^{N-1}(N-n)\hat{Q}_n^2 + (N-1)\varepsilon_0^2 \\
     &+ 2\sum_{n=2}^{N-1}\sum_{j=1}^{n-1}(N-n)\hat{Q}_j\hat{Q}_n  \\
     &+ 2\varepsilon_0\sum_{n=1}^{N-1}(N-n)\hat{Q}_n.
 \end{split}
 \end{equation}
 
 In order to eliminate the gauge fields operators $\hat{U}_n$ and $\hat{E}_n$ from Eq.~\eqref{eq:higgs_1d}, we can apply a residual gauge transformation analogously to the procedure presented in \cite{muschik_u1_2017}, which yields
 \begin{subequations}
     \begin{align}
     \hat{\phi}_n &\rightarrow \left(\prod_{k=1}^{n-1}\hat{U}_k^\dagger\right) \hat{\phi}_n, \\
     \hat{\phi}_n^\dagger &\rightarrow \hat{\phi}_n^\dagger \left(\prod_{k=1}^{n-1}\hat{U}_{n-k}\right).
     \end{align}
 \end{subequations}
 This induces the transformation,
 \begin{equation}
     \hat{\phi}_n^\dagger \hat{U}_n^\dagger \hat{\phi}_{n+1} \rightarrow \hat{\phi}_n^\dagger \hat{\phi}_{n+1},
 \end{equation}
 which in combination with Eq.~\eqref{eq:e_der} may be applied to Eq.~\eqref{eq:higgs_1d} and subsequently results in the effective Hamiltonian given in Eq.~\eqref{eq:higgs_e}. Here, the gauge degrees of freedom have been eliminated completely, which came at the cost of introducing long-range interactions between the Higgs fields.
A method for obtaining the effective Hamiltonian in more than one spatial dimension is described in Refs.~\cite{Paulson:2020zjd, Haase:2020kaj}.

 \section{Generation of $\hat{H}_{NN}$ interactions in a parametric cavity \label{app:4op_der}}

We now describe how the interaction term $\hat{H}_{NN}$ in Eq.~\eqref{eq:4op_inter}, which is one of the resource Hamiltonians used in the VQE, can be generated on the microwave platform. The derivation is analogous to the one presented in Ref.~\cite{sandbo_thesis}. Using a symmetric SQUID, the fourth-order term in the SQUID cosine potential is given by
\begin{equation}
\hat{H}_{\text{SQ}} = g_0\left(\hat{a}_p + \hat{a}_p^\dagger\right)\left(\sum_{n=1}^N\hat{a}_n + \hat{a}_n^\dagger\right)^4,
\end{equation}
where $\hat{a}_p$ ($\hat{a}_n$) is the annihilation operator for the pump mode (cavity mode $n$) and $g_0$ is the intrinsic interaction strength between the pump and the cavity modes. The higher-order terms in the cosine potential are neglected. As a result, the full Hamiltonian of the superconducting cavity reads
\begin{subequations}
	\begin{align}
	&\hat{H} = \hat{H}_0 + \hat{H}_{\text{SQ}}, \\
	&\hat{H}_0 = \omega_p\hat{N}_p + \sum_{n=1}^N\omega_n\hat{N}_n,
	\end{align}
\end{subequations}
where $\omega_p$ is the natural frequency of the pump and $\omega_n$ is the natural frequency of mode $n$. We then transform $\hat{H}$ into the interaction picture w.r.t $\hat{H}_0' = \omega_p\hat{N}_p + \sum_n \omega_n'\hat{N}_n$, which yields the Hamiltonian
\begin{equation}
\begin{split}
&\hat{H}_{\text{int}} = \sum_{n=1}^N(\omega_n - \omega_n')\hat{N}_n \\
&+ g_0\left(\hat{a}_pe^{-i\omega_pt} + \text{\text{H.C.}}\right) \left(\sum_{n=1}^N\hat{a}_ne^{-i\omega_n't} + \text{\text{H.C.}} \right)^4.
\end{split}\label{eq:h_int}
\end{equation}
The first term of $\hat{H}_\text{int}$ can be identified with the free rotation part of the unitary evolution given in Eq.~\eqref{eq:4op_inter}, with $\Omega_n = \omega_n-\omega_n'$. 
For the remainder of this derivation we now focus on the second term in Eq.~\eqref{eq:h_int}. The expansion of the fourth power factor yields a number of terms; as shown in Ref.~\cite{sandbo_thesis}, a suitable choice of the pump mode can selectively enhance the desired interaction among the terms available in the expansion. We are interested in generating interactions between photon number operators of different modes, which are terms that are time independent in the expansion. By choosing $\omega_{p}$ much smaller than any other frequency that appears in the fourth power expansion, we can use the rotating wave approximation to neglect all terms that rotate at frequency larger than $\omega_{p}$. We assume that $\omega_{p}$ is so small that it can be effectively considered zero for the duration of the experiment. Let us now calculate the effective Hamiltonian explicitly. There are two terms in the expansion that are time independent: a product of $\{\hat{a}^\dagger_i, \hat{a}_i, \hat{a}^\dagger_j, \hat{a}_j\}$, and a product of $\{\hat{a}^\dagger_i, \hat{a}_i, \hat{a}^\dagger_i, \hat{a}_i\}$. Using combinatorics and the bosonic commutation relations, the first type of terms sum to
\begin{equation}
\label{eq:comb1}
24\sum_{n=2}^N\sum_{k=1}^{n-1}\hat{N}_k\hat{N}_n + 12(N-1)\sum_{n=1}^N\hat{N}_n + 3N(N-1)
\end{equation}
and the second type of terms sum to
\begin{equation}
\label{eq:comb2}
6\sum_{n=1}^N\hat{N}^2_n + 6\sum_{n=1}^N\hat{N}_n + 3N.
\end{equation}
Combining Eqs.~(\ref{eq:comb1}-\ref{eq:comb2}) together gives Eq.~\eqref{eq:4op_inter} [the constant term is ignored since it contributes an overall phase to Eq.~\eqref{eq:4op_inter}]. Finally, we assume that the pump has a strong coherent tone such that we can apply the parametric approximation and substitute $\hat{a}_p$ with its classical amplitude $|\alpha|e^{i\phi}$. With this approximation the time evolution induced by  $\hat{H}_\text{int}$ is exactly the one given in Eq.~\eqref{eq:4op_inter} with  $g\prime = 2g_0|\alpha|\cos{\phi}$. 

\section{Truncation effects in the HOBM mapping \label{app:truncation}}
 In this section, we consider the effects of using a truncated HOBM mapping (given in Sec.~\ref{sec:HOBM}), for our VQE simulations. In the following we omit the lattice site index as these expressions hold for all sites. Using the untruncated HOBM model, all of the commutation relations remain the same (see Sec.~\ref{subsec:Hamiltonian} for the commutation relations of the original model), except for
\begin{align} 
\label{eq:untruncated}
[\hat{\phi}, \hat{\phi}^\dagger] \rightarrow  \frac{1}{N_0}.
\end{align}
The correct commutation relation is $[\hat{\phi}, \hat{\phi}^\dagger] = 0$, and it is recovered as $N_0\to \infty$. Let $\hat{a}(k)$ and $\hat{N}(k)$ be the truncated operators of size $2k+1$.
When we consider the truncated HOBM model (as is the case for our classical simulation) an error is introduced, and the following commutation relations can be derived
\begin{subequations}
	\begin{equation}
	\begin{split}
	[\hat{Q}, \hat{\phi}^\dagger] \rightarrow& \left[\hat{N}(k) - N_0, \frac{1}{\sqrt{N_0}}\hat{a}^\dagger(k)\right] \\
	=& \frac{1}{\sqrt{N_0}}\hat{a}^\dagger(k) \\ &+ \frac{N_0-k}{\sqrt{N_0}}\hat{a}^\dagger(k)\ket{N_0-k}\bra{N_0-k}, \label{eq:comm_hobm1}
	\end{split}
	\end{equation}
	\begin{equation}
	\begin{split}
	[\hat{\phi}, \hat{\phi}^\dagger] \rightarrow& \frac{1}{N_0}[\hat{a}(k), \hat{a}^\dagger(k)] \\
	=&\frac{1}{N_0} +\frac{N_0-k}{N_0}\ket{N_0-k}\bra{N_0-k} \\
	&- \frac{N_0+k+1}{N_0}\ket{N_0+k}\bra{N_0+k}. \label{eq:comm_hobm2} 
	\end{split}
	\end{equation}
\end{subequations}
To recover the correct commutation relation in Eq.~\eqref{eq:comm_hobm1}, we set $k = N_0$, which gives
\begin{subequations}
	\begin{equation} 
	[\hat{Q}, \hat{\phi}^\dagger] \rightarrow \left[\hat{N}(k)-N_0, \frac{1}{\sqrt{N_0}}\hat{a}^\dagger(k)\right] =  \frac{1}{\sqrt{N_0}}\hat{a}^\dagger(k), \label{eq:comm1}
	\end{equation}
	\begin{equation}
	\begin{split}
	[\hat{\phi}, \hat{\phi}^\dagger] &\rightarrow \frac{1}{N_0}\left[\hat{a}(k), \hat{a}^\dagger(k)\right] \\
	&=  \frac{1}{N_0} - \frac{2N_0+1}{N_0}\ket{2N_0}\bra{2N_0}.  \label{eq:comm2}
	\end{split}
	\end{equation}
\end{subequations}
Eq.~\eqref{eq:comm1} is the same commutation relation as in the full Hilbert space [Eq.~\eqref{eq:first_comm}] and, given that $[\hat{\phi}, \hat{\phi}^\dagger] = 0$ in the full Hilbert space, the truncation error in Eq.~\eqref{eq:comm2} goes to zero as $N_0 \to \infty$. In general, Eq.~\eqref{eq:comm2} shows that the state with maximum photon number, $\ket{2N_0}$ is most affected by truncation errors.

 \section{Higgs mechanism and phase structure in the absence of a topological term \label{app:higgs_mechanism}}
 
 In this appendix we briefly review the Brout-Englert-Higgs mechanism for the continuum model and discuss its implications for the phase structure. To this end let us start from the continuum Lagrangian
 \begin{align}
     \mathcal{L} = (D_\mu \varphi)^* (D_\mu \varphi)  - \frac{1}{4}F_{\mu\nu}F^{\mu\nu} - V(|\varphi|),
     \label{eq:higgs_lagrangian}
 \end{align}
 where $\varphi$ is a classical complex scalar field, $D_\mu = \partial_\mu + igA_\mu$ the covariant derivative with the gauge field $A_\mu$ and the coupling strength $g$, $F_{\mu\nu} = \partial_{\mu}A_{\nu} - \partial_{\nu}A_{\mu}$ the field strength tensor and $V(|\varphi|) = -m^2|\varphi|^2 + \frac{\lambda}{2}|\varphi|^4$ with $\lambda > 0$. The first term of the action describes the kinetic energy of the scalar field and the coupling to the gauge field, the second term the kinetic energy of the gauge field, and the potential $V(|\varphi|)$ contains the mass term and the self-interaction of the scalar field. It is straightforward to see that the action in Eq.~\eqref{eq:higgs_lagrangian} is invariant under U(1) gauge transformations given by
 \begin{align*}
     \varphi(x) \to e^{ig\alpha(x)}\varphi(x),\quad\quad A_\mu(x) \to A_\mu - \partial_\mu \alpha(x),
 \end{align*}
 where $\alpha(x)$ is a real, differentiable function.
 
 To get an intuition about the physics of the model, it is instructive to derive a semiclassical picture. To this end, we assume that the potential term is dominant and in the ground state the field $\varphi$ fluctuates only slightly around the vacuum expectation value $\varphi_0$ minimizing the potential $V(|\varphi|)$. Rewriting the potential as
 \begin{align*}
     V(|\varphi|) = \frac{\lambda}{2}\left[\left(|\varphi|^2- \frac{2m^2}{\lambda}\right)^2 - \left(\frac{m^2}{\lambda} \right)^2\right],
 \end{align*}
 we see that $V(|\varphi|)$ is quadratic function of $|\varphi|^2$ and we can easily read off the minimum. Taking into account that $|\varphi|^2$ is a positive semi-definite quantity, we have to distinguish two cases depending on the sign of $m^2$. (i) For $m^2 \leq 0$ the potential is a parabola in $|\varphi|^2$ (see also lower left inset of Fig.~\ref{fig:pd_higgs_no_theta}) with a unique minimum at $\varphi_0 = 0$. For $\varphi\approx 0 $ Eq.~\eqref{eq:higgs_lagrangian} reduces to
 \begin{align*}
     \mathcal{L} \approx -\frac{1}{4}F_{\mu\nu}F^{\mu\nu},
 \end{align*}
 which is nothing but a \emph{pure gauge theory} describing a \emph{massless photon} and showing \emph{charge confinement}~\cite{Heitger1997}. (ii) For $m^2 > 0 $ the potential has the shape of a ``Mexican hat'' (see also upper right inset of Fig.~\ref{fig:pd_higgs_no_theta}) with the minima forming a level set given by 
 \begin{align}
     |\varphi_0|^2 = \sqrt{\frac{m^2}{\lambda}} =: \frac{v}{\sqrt{2}}.
     \label{eq:higgs_minima}
 \end{align}
 In the expression above, we have defined the quantity $v = \sqrt{2m^2/\lambda}$ for convenience. To account for quantum fluctuations around $\varphi_0$, we parameterize the field using polar representation
 \begin{align*}
     \varphi(x) = \frac{v + h(x)}{\sqrt{2}} e^{i\frac{\phi(x)}{v}},
 \end{align*}
 where the real fields $h(x)$ and $\phi(x)$ describe the fluctuations of the length of the field in radial direction and the phase. Combining this expression with Eq.~\eqref{eq:higgs_lagrangian} and considering only at the gauge part $\mathcal{L}_\mathrm{gauge}$ of the resulting Lagrangian we find
 \begin{align}
     \mathcal{L}_\mathrm{gauge} = - \frac{1}{4}F_{\mu\nu}F^{\mu\nu} + \frac{1}{2} m_p^2A_\mu A^\mu,
     \label{eq:higgs_lagrangian_massive}
 \end{align}
 showing that the \emph{photon is massive} with mass $m_p = gv$. Moreover, in that case the \emph{U(1) symmetry is spontaneously broken} as the vacuum of the theory corresponds to a single one of the minima described by Eq.~\eqref{eq:higgs_minima}. Sending $\lambda\to\infty$ while keeping the ratio $m^2/\lambda$ fixed, the radial fluctuations can be neglected and $\varphi$ has a fixed length of $v/\sqrt{2}$. Inserting this expression into Eq.~\eqref{eq:higgs_lagrangian} one finds the effective Lagrangian
 \begin{align}
 \begin{split}
     \mathcal{L}_\mathrm{eff} =& \frac{1}{2}\partial_\mu \phi\partial^\mu \phi + \frac{1}{2} m_p^2 A_\mu A^\mu \\&+ m_p^2 A_\mu\partial^\mu\phi  - \frac{1}{4}F_{\mu\nu}F^{\mu\nu}.
     \label{eq:higgs_lagrangian_effective}
\end{split}
 \end{align}
 From Eq.~\eqref{eq:higgs_lagrangian_effective} one can derive the Hamiltonian in Eq.~\eqref{eq:higgs_1d} as shown in Ref.~\cite{Gonzalez-Cuadra2017}.
 
 The simple semiclassical picture above shows that the model has two distinct regions. In the Higgs region the U(1) symmetry of the theory is spontaneously broken and the photon acquires a mass. In contrast, the confining region corresponds to a pure gauge theory describing a massless photon and the U(1) symmetry is intact. Going beyond this simple semiclassical picture and solving the the lattice discretization of Eq.~\eqref{eq:higgs_lagrangian} numerically using MCMC methods one finds that the intuition from the rather simple semiclassical picture also holds true more generally~\cite{Jones1979,Heitger1997} and one obtains the phase diagram shown in Fig.~\ref{fig:pd_higgs_no_theta} in the main text.
 
 \section{Periodicity and Symmetries of the Hamiltonian \label{app:symmetries}}
 
 Here we briefly show that physics is periodic in $\ez$ with period $1$ and discuss the symmetries of the lattice Hamiltonian. For simplicity we work with the formulation in Eq.~\eqref{eq:higgs_1d}, in which the gauge field has not been integrated out yet. 
 
 To show that physics is periodic in $\ez$, we consider the transformation
 \begin{equation}
     \begin{split}
     \hat{\phi}_n &\to \hat{\phi}_n^\dagger, \\ \hat{Q}_n&\to -\hat{Q}_n,\\
     \hat{U}_n &\to \hat{U}_n^\dagger, \\ \hat{E}_n&\to -(\hat{E}_n + k),\quad k\in\mathds{Z}.
     \end{split}
 \label{eq:generalized_charge_conjugation}
 \end{equation}
 Let us first focus on $k=0$. In that case Eq.~\eqref{eq:generalized_charge_conjugation} corresponds to a charge conjugation transformation $\hat{\mathcal{C}}_n$ which exchanges particles and antiparticles. This transformation is unitary, and, in particular, we see that applying it twice we get the initial operators back, thus showing that $\hat{\mathcal{C}}_n^2 =1$ and charge conjugation is a $\mathds{Z}_2$ symmetry. 
 
 To get further insight into the case $k\neq 0$, we look at the commutation relation in Eq.~\eqref{eq:electric_field_commutation}, from which follows that the unitary operator $\hat{U}_n$ introduces integer shifts in $\hat{E}_n$:
 \begin{equation*}
     \hat{U}^\dagger_n \hat{E}_n \hat{U}_n = \hat{E}_n -1,\quad\quad \hat{U}_n \hat{E}_n \hat{U}^\dagger_n = \hat{E}_n + 1.
 \end{equation*}
 Hence, by combining charge conjugation with $\hat{U}_n$ to $\hat{U}^k_n \hat{\mathcal{C}}_n$ [$(\hat{U}^\dagger_n)^k \hat{\mathcal{C}}_n$], we obtain an additional shift in the electric field by $k$ positive (negative) units. Since both $\hat{\mathcal{C}}_n$ and $\hat{U}_n$ are unitary, Eq.~\eqref{eq:generalized_charge_conjugation} indeed describes a unitary transformation. Applying this unitary transformation to Gauss's law in Eq.~\eqref{eq:gauss_law2} and the Hamiltonian in Eq.~\eqref{eq:higgs_1d}, we see that all terms are invariant except for the electric field energy which transforms according to
 \begin{align*}
 \begin{split}
     \sum_n (\ez + \hat{E}_n)^2 \to& \sum_n (\ez -\hat{E}_n - k )^2 \\=& \sum_n (\hat{E}_n + k - \ez)^2.
\end{split}
 \end{align*}
 Looking at that equation, we make the following observations. (i) For any value of $\ez$ we can find a $k$ such that the Hamiltonian is mapped to a unitarily equivalent one with $k - \ez\in[0,1)$. Consequently, physics is periodic in $\ez$ with period $1$, and we can restrict ourselves to $\ez \in [0,1)$ without loss of generality. (ii) For $k=0$, $\ez=0$ and $k=1$, $\ez=1/2$ the transformation from Eq.~\eqref{eq:generalized_charge_conjugation} is a symmetry of the Hamiltonian. In particular, for $\ez=1/2$ there are two field configurations yielding the same electric energy. Thus, for $R^2\to 0$ the ground state of the Hamiltonian is doubly degenerate and has a $\mathds{Z}_2$ symmetry. This symmetry is preserved for non-vanishing $R^2$ along the critical line, before it is eventually spontaneously broken upon reaching the critical value $R_c^2$.
 
 \section{Location of the first phase transition \label{app:spike_der}}
 
 In this appendix we derive Eq.~\eqref{eq:jump_formula} from Sec.~\ref{subsec:phase_with_topological_term}. This formula gives the smallest, positive value of $\varepsilon_0$ where a phase transition occurs and the dominant term in the ground state after the phase transition. Assuming that $R^2$ is small such that the kinetic term in Eq.~\eqref{eq:higgs_e} can be ignored, the Hamiltonian (including the penalty term for vanishing total charge) becomes
 \begin{equation}
     \begin{split}
     \hat{H} &= \sum_{n=1}^N \left(\left(\frac{1}{2R^2} + \frac{N-n}{2\beta} + \ell\right)\hat{Q}_{n} + \frac{\varepsilon_0(N-n)}{\beta}\right)\hat{Q}_n \\
     &+ \sum_{n=2}^{N}\sum_{j=1}^{n-1}\left(\frac{N-n}{\beta} + 2\ell\right) \hat{Q}_j\hat{Q}_n \\
     &- \beta(N-1) + \frac{\varepsilon_0^2}{2\beta}(N-1). \label{eq:jump_ham}
     \end{split}
 \end{equation}
 Looking at Eq.~\eqref{eq:jump_ham}, we see that the eigenstates of $\hat{H}$ in this approximation are given by charge eigenstates. Before the transition occurs, meaning in the regime of small $\varepsilon_0$ and $R^2$, we expect the ground state to be dominated by the state with vanishing charge everywhere, $\ket{0}^{\otimes N}$. The corresponding energy is given by $(N-1)( \frac{\varepsilon_0^2}{2\beta} -\beta)$. 
Since we work in the subsector of vanishing total charge and for $R^2 \ll 1$ the $\sum_n c_n\hat{Q}_n^2$ term in Eq.~\eqref{eq:jump_ham} strongly penalizes states with many nonzero charges, we expect the ground state after the phase transition to be of the form  $\ket{\psi} = \ket{0}^{\otimes n_1-1}\ket{k}\ket{0}^{\otimes n_2-n_1-1}\ket{-k}\ket{0}^{\otimes N-n_2}$ with $n_2 > n_1$ and $k \in \mathds{Z}$. The energy of this state is given by
 \begin{equation}
     \frac{k^2}{R^2} + \frac{k^2r}{2\beta} + \frac{k\varepsilon_0r}{\beta}
     -\beta(N-1) + \frac{\varepsilon_0^2}{2\beta}(N-1),
 \end{equation}
 where $r = n_2-n_1$. Since the ground-state energy is a continuous function of $\ez$, we can equate this expression to the energy before the transition which gives
 \begin{equation}
     \varepsilon_0 = -k\left(\frac{1}{2} + \frac{\beta}{R^2r}\right).
 \end{equation}
 Since $\varepsilon_0>0$, that forces $k<0$. So, for each value of $k$ and $r$, this formula predicts where the first phase transition could occur. Of course, of all the possible choices of $k$ and $r$, the one that predicts the smallest value for the phase transition point $\varepsilon_0$ gives the correct formula. Since $k \neq 0$ (no phase transition would occur in that case), we therefore choose $k = -1$ and $r = N-1$. Thus, we predict the first phase transition to cause the ground state to transform from $\ket{0}^{\otimes N}$ to  $\ket{-1}\ket{0}^{\otimes N-2}\ket{+1}$ at the point
 \begin{equation}
     \varepsilon_0 = \frac{1}{2} + \frac{\beta}{R^2(N-1)}.
 \end{equation}

 \section{General VQE protocol}
 \label{app:general_vqe}
 
 In this appendix we provide the VQE strategy used to obtain Fig.~\ref{fig:Rsq1}. Approximating the ground state of the model in the vicinity of the critical point $R^2 \approx R^2_c$ requires a more complex and deeper variational circuit than the one used for detecting the first-order phase transition for smaller values of $R^2$ in Fig.~\ref{fig:Rsq03}. Furthermore, the number of variational parameters quickly becomes unfeasible for the type of Bayesian optimizer we considered \cite{Acerbi2017Practical, Frazier2018A-Tutorial}, therefore we need to adopt a strategy where only one part of the circuit is optimized at a time.
 
 The basic element of the circuit is a combination of beam splitters connecting all possible sites, alternated with the interactions in $\hat{H}_{NN}$ [see Eq.~\eqref{eq:4op_inter}]
 \begin{equation}
 	\hat{D}^{k} (\vec{\Omega},\vec{\theta}^{k})= \prod_{i = 1}^{N-1} \prod_{j = 1+1}^{N} \hat{C}_{(i,j)}(\vec{\Omega},\vec{\theta}^{k}),
 \end{equation}
 where $\hat{C}_{(i,j)}$ is a generalization of the operator given in Eq.~\eqref{eq:C14}, obtained by choosing $N_{l} = 1$, and using the beam-splitter interaction acting on modes $(i,j)$.
 
 The unitary $\hat{D}^{k}$ is then repeated for a suitable number of layers such that the number of variational parameters is feasible for the chosen optimization algorithm to obtain the block
 \begin{equation}
 	\hat{B} = \prod_{k = 1}^{N_l} \hat{D}^{k}(\vec{\Omega},\vec{\theta}^{k}).
 \end{equation}
 
 The optimization proceeds as follows. To the initial state $\ket{1223}$ we apply the unitary $B$. For this block,  both the rotations $\vec{\Omega}$ and the set of $\{\vec{\theta}^{k}\}$ are considered as variational parameters. After the convergence of this first optimization we iteratively apply further unitaries $\hat{B}$ to the previously obtained state and optimize the parameters $\{\vec{\theta}^{k}\}$ associated to each new block. The rotations $\vec{\Omega}$ are chosen by going into a suitable frame of reference and therefore have to be consistent throughout the variational circuit, hence they are only considered in the first block. To obtain the results show in Fig.~\ref{fig:Rsq1} we found that the optimization in the case of the VQE without including statistical noise was successful by choosing $N_{l}= 2$ for the first block and  $N_{l}= 3$ for each subsequent block. In total, we employed five blocks to obtain the results shown in the main text.

\bibliographystyle{quantum}

 \end{document}